\documentclass[11pt,numberedappendix,twocolappendix,]{emulateapj}

\usepackage{natbib}
\usepackage{amsmath}
\usepackage{amssymb}
\usepackage{aas_macros}
\usepackage{appendix}
\usepackage{multirow}
\usepackage{color,xcolor}
\usepackage{url}
\usepackage{ulem}
\usepackage{hyperref}
\usepackage{textcomp}
\hypersetup{colorlinks,linkcolor={blue!50!black},citecolor={blue!50!black},urlcolor={blue!50!black}}

\def\water{H$_2$O}
\def\xwater{$x_\text{\water}$}
\def\preslevel{1.39 mbar}
\def\modelE{ROCKE-3D}
\def\wv{water vapor}

\shorttitle{NIR-Driven Moist Upper Atmospheres of Temperate Terrestrial Exoplanets}
\shortauthors{Fujii et al.}

\begin{document}


\title{NIR-Driven Moist Upper Atmospheres of Synchronously Rotating Temperate Terrestrial Exoplanets}

\author{Yuka Fujii}
\affil{NASA Goddard Institute for Space Studies, 2880 Broadway, New York, NY, USA}
\affil{Earth-Life Science Institute, Tokyo Institute of Technology, Ookayama, Meguro, Tokyo 152-8550, Japan}

\author{Anthony D. Del Genio}
\affil{NASA Goddard Institute for Space Studies, 2880 Broadway, New York, NY, USA}

\author{David S. Amundsen}
\affil{NASA Goddard Institute for Space Studies, 2880 Broadway, New York, NY, USA}
\affil{Department of Applied Physics and Applied Mathematics, Columbia University, New York, NY 10025, USA}


\begin{abstract}
H$_2$O is a key molecule in characterizing atmospheres of temperate terrestrial planets, and observations of transmission spectra are expected to play a primary role in detecting its signatures in the near future.  
The detectability of H$_2$O absorption features in transmission spectra depends on the abundance of water vapor in the upper part of the atmosphere. 
We study the three-dimensional distribution of atmospheric H$_2$O for synchronously rotating Earth-sized aquaplanets using the general circulation model (GCM) ROCKE-3D, and examine the effects of total incident flux and stellar spectral type.  
We observe a more gentle increase of the water vapor mixing ratio in response to increased incident flux than one-dimensional models suggest, in qualitative agreement with the climate-stabilizing effect of clouds around the substellar point previously observed in GCMs applied to synchronously rotating planets. 
However, the water vapor mixing ratio in the upper atmosphere starts to increase while the surface temperature is still moderate. 
This is explained by the circulation in the upper atmosphere being driven by the radiative heating due to absorption by water vapor and cloud particles, causing efficient vertical transport of water vapor. 
Consistently, the water vapor mixing ratio is found to be well-correlated with the near-infrared portion of the incident flux. 
We also simulate transmission spectra based on the GCM outputs, and show that for the more highly irradiated planets, the H$_2$O signatures may be strengthened by a factor of a few, loosening the observational demands for a H$_2$O detection.
\end{abstract}



\section{Introduction}
\label{s:intro}

One of the primary interests in future observations of Earth-sized planets is the presence of \water{}, both as a tracer of habitable conditions and as a clue to the origin and evolutional pathways of the planets. 
Its spectral signatures may be targeted in the near future through transmission spectroscopy using JWST or next-generation ground-based telescopes.
Although \water{} signatures have been detected in the atmospheres of hot Jupiters \citep[e.g.][]{Tinetti2007,Sing2016}, detecting molecular signatures, including \water{}, on temperate terrestrial planets is exceedingly challenging \citep{Cowan2015} because of the small planetary radius and the small scale height (due to the lower temperature and presumably larger mean molecular weight). 
Additionally, while \water{} is a dominant opacity source of Earth's atmosphere, its signatures in the transmission spectra of temperate planets are not strong due to the ``cold trap'', i.e., \wv{} evaporated from the surface is transported upward, but condenses as air rises and cools, and most of it precipitates, leaving the stratosphere---where transmission spectroscopy typically probes---fairly dry. 
Several studies modeling transmission spectra of the Earth found modest signatures of \wv{} \citep[e.g.][]{Ehrenreich2006, Kaltenegger2009, Betremieux2013, Misra2014}. 
However, the efficiency of the cold trap depends on the atmospheric properties in general. 
Thus, modeling the \wv{} mixing ratio profiles is an issue of interest in the context of evaluating the observability of atmospheric signatures of transiting terrestrial planets. 

The mixing ratio of \wv{} in the stratosphere is also closely related to planetary water loss and thus planetary habitability. 
\citet{Kasting1993} considered the maximum stratospheric \wv{} mixing ratio that would allow for a planet to sustain an amount of water equivalent to the  Earth's oceans over the Earth's age, $\sim 3 \times 10^{-3}$, as a criterion for the inner edge of the habitable zone (the threshold for the moist-greenhouse stage). 
Motivated mainly by this ``habitability'' aspect, atmospheric models with different levels of complexity have examined stratospheric \wv{}, or equivalently, the efficiency of the cold trap, with varying planetary properties. 

\citet{Kasting1993} and \citet{Kopparapu2013} used a simple one-dimensional (1D) model with a saturated troposphere and an isothermal stratosphere to examine the response of the stratospheric \wv{} mixing ratio to increased instellation \citep[see also][]{Kasting2015}. 
They found a transition from an Earth-like dry upper atmosphere to a water-dominated atmosphere, driven by the increase of the surface temperature. 
The above-mentioned critical mixing ratio ($\sim 3 \times 10^{-3}$) is reached when the surface temperature is $\sim $340 K. 
\citet{Kodama2015} also showed a similarly rapid transition to a water-dominated atmosphere as a function of incident flux using a 1D radiative-convective model.  

\citet{Wordsworth2013,Wordsworth2014} argued that the degree to which a condensible gas (e.g., water) is transported from the surface is represented by $\mathcal{M} = m_v p_v L / m_n p_n c_p T $ where $L$, $c_p$, and $T$ are the specific latent heat of the condensing gas, the specific heat capacity at constant pressure of the non-condensing gas, and surface  temperature, respectively, while $m_{\{v,n\}}$ and $p_{\{v,n\}}$ are molar masses and the partial pressures of the  condensing ($v$) or non-condensing ($n$) gases. 
A moist upper atmosphere occurs when $\mathcal{M} > 1$. 
This indicates that water vapor profiles also depend on the background atmospheric properties, in addition to incident flux. 

Atmospheric profiles of template terrestrial planets have also been studied using photochemical models combined with 1D radiative-convective models \citep{Segura2003,Segura2005,Rauer2011,Hedelt2013,Rugheimer2013,Rugheimer2015}. 
\citet{Rauer2011} and \citet{Rugheimer2013,Rugheimer2015} observed an enhanced water mixing ratio in the stratosphere for planets around late-type stars, which was attributed to the higher surface temperatures and photochemical production from CH$_4$. 

Inevitably, 1D models rely on several assumptions that cannot be treated within the model, including the global transport of heat and water vapor, as well as the effect of clouds. 
Various three-dimensional (3D) general circulation models (GCMs) revisited the atmospheric structure of terrestrial planets and have demonstrated the importance of accounting for the 3D circulation and climate heterogeneity \citep[e.g.][]{Ishiwatari2002,Abe2011,Leconte2013a,Leconte2013b,Wolf2014,Wolf2015}. 
Among others, \citet{Yang2013} pointed out that for synchronously rotating planets, the convection around the substellar point produces optically thick clouds that increase the planetary albedo and contribute to keep the surface temperature moderate for a much wider range of instellation \citep[see also][]{Yang2014,Way2015,Kopparapu2016}. 
This situation is of particular relevance to the planets in the habitable zones of late-type stars, as they are likely to be synchronously rotating due to tidal locking \citep{Dole1964, Kasting1993}, although this need not always occur \citep{Goldreich1966,Leconte2015}. 
They primarily studied the highest instellation that allows for the model to converge as a proxy for the inner edge. 
While they also presented the stratospheric water vapor amounts, the discussion was limited. 

Motivated by these works on the stabilized climate of synchronously rotating planets, we are specifically interested in the \wv{} transport into the upper atmosphere, not only because it affects the water loss rate (and thus the conventional criterion for habitable conditions), but also because it limits our ability to detect \water{} signatures of planets with surface liquid water, through transmission spectroscopy. 
Indeed, planets around later-type stars, which are more likely to be synchronously rotating, are best suited for future transit observations for several reasons, including the higher transit probability, larger signals in transmission spectra, and the abundance of late-type stars in the nearby universe. 

In order to obtain better insights into the processes that control the \wv{}  mixing ratio in the upper atmosphere of synchronously rotating terrestrial planets with surface water, we conduct a series of experiments using the \modelE{} GCM, which calculates the atmospheric dynamic, thermodynamic, and radiative fields up to $0.1$ mbar ($\sim $65 km in the case of the Earth), sufficiently resolving the stratosphere. 
We specifically study the dependence of the \wv{} mixing ratio on the incident flux (total incident flux and the spectral type), fixing planetary parameters (radius, gravity, atmospheric composition, and atmospheric pressure). 
We also present synthetic transmission spectra based on GCM outputs in order to evaluate the detectability of \water{} signatures with future transmission spectroscopy. 

The paper is organized as follows: %
Section \ref{s:model} describes the setup of our GCM experiments and  calculations of transmission spectra. %
Section \ref{s:results} discusses our results of \wv{} profiles as a function of the incident spectrum, and shows the modeled transmission spectra. %
The sensitivities of our results to the ocean model, the CO$_2$ abundance, the spin period, and the water vapor continuum absorption are studied in Section \ref{s:sensitivity}. %
Finally, Section \ref{s:summary} summarizes our findings. %

\section{Model}
\label{s:model}

\subsection{GCM experiments}

We use the \modelE{} GCM \citep{Way2017} to obtain 3D atmospheric structures of terrestrial planets with surface water under varying irradiation, focusing on the \water{} profile in the upper atmosphere. 
\modelE{} is a generalization of the ModelE2 GCM \citep{Schmidt2014}, which has been developed for the Earth at NASA's Goddard Institute for Space Studies. 
The details of the ROCKE-3D GCM are presented in \citet{Way2017}.

In the following, we first describe the specific configurations in our experiments to clarify our targets. Then, we also discuss in some detail the convection/cloud parameterizations and the radiation parameterization used in our GCM, as these are of particular importance to our results. 

\begin{table*}[!btp]
\caption{Stellar Properties}
\begin{center}
\begin{tabular}{ccllllrr} \hline \hline
Star & Type & $T_{\rm eff}$ & $L$~[$L_{\odot}$] & $M$~[$M_{\odot}$] & $R$~[$R_{\odot}$] & reference & source of spectrum \\ \hline
Sun & G2V & 5772~K & 1 & 1 & 1 & - & \citet{Lean2005}, SPARC/SOLARIS\footnotemark[1] \\ 
HD 22049 & K2V & 5039~K & 0.32 & 0.82 & 0.74 & \citet{Baines2012} & \citet{Segura2003} \\
Kepler-186 & M1V & 3755~K & 0.055 & 0.544 & 0.523 & \citet{Torres2015} & \citet{Allard2012} \\
GJ 876 & M4V & 3129~K & 0.0122 & 0.37 & 0.3761 & \citet{vonBraun2014} & \citet{Domagal-Goldman2014} \\ \hline
\end{tabular}
\end{center}
\label{tbl:stellar_properties}
\footnotetext[1]{\url{http://solarisheppa.geomar.de/ccmi}}
\end{table*}

\begin{table}[btp]
\caption{Orbital period, to which the rotation period is equated, in the case of $S_X=1$. }
\begin{center}
\begin{tabular}{ccr} \hline \hline
Star & $T_{\rm eff}$ & Period [days] ($S_X=1$) \\ \hline
Sun & 5772~K & 365.3 \\ 
HD 22049 & 5039~K & 171.6 \\
Kepler 186 & 3755~K & 56.2 \\
GJ 876 & 3129~K & 22.0 \\ \hline
\end{tabular}
\end{center}
\label{tbl:orbital_period}
\end{table}

\begin{figure}[!b]
    \begin{center}
    \includegraphics[width=\hsize]{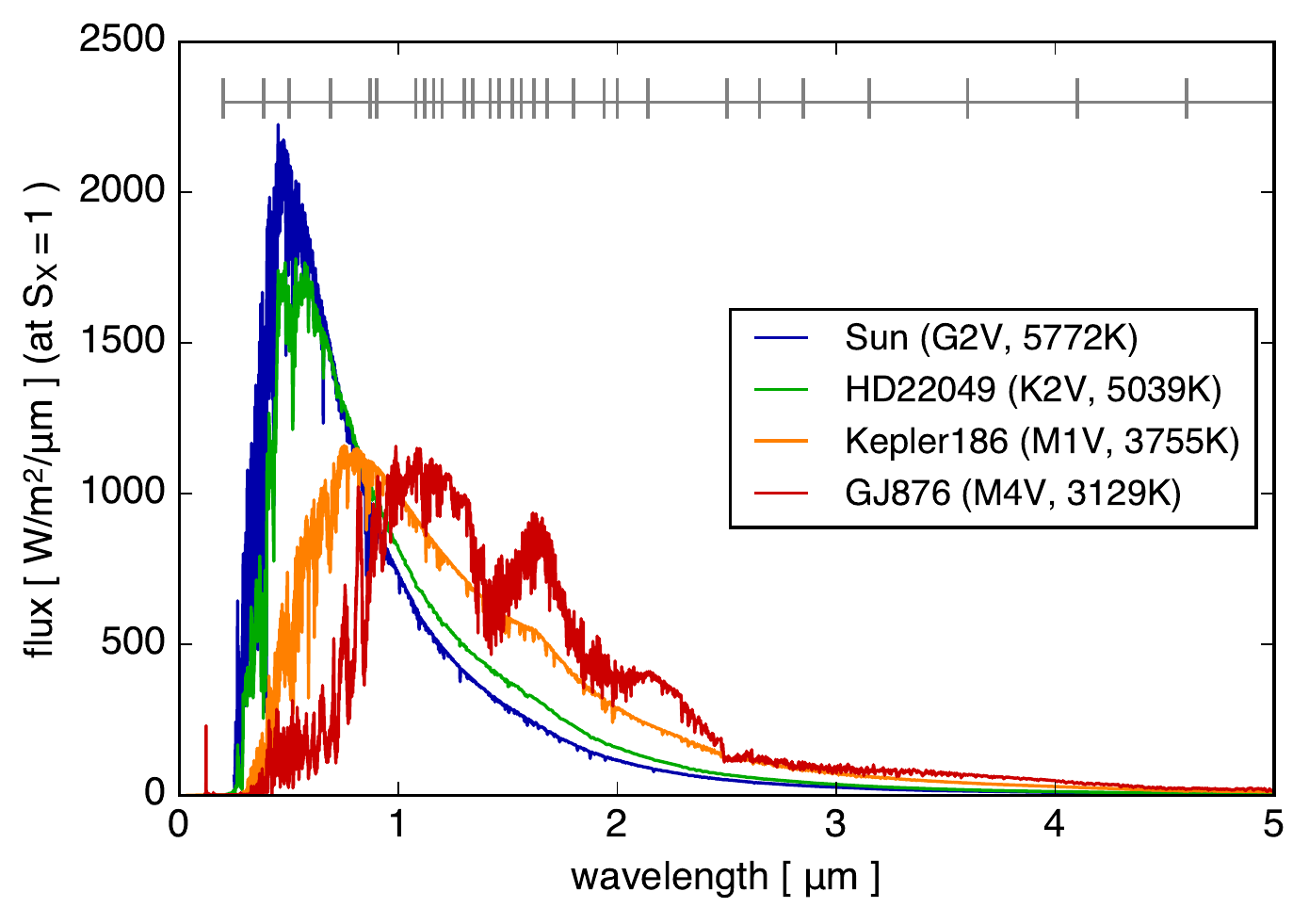}
    \end{center}
\caption{Spectra of the stars considered in this paper. The flux is normalized so that the total incident flux matches the solar constant. The data sources are listed in the rightmost column of Table~\ref{tbl:stellar_properties}. The resolution of the spectra was lowered from the original data just for display purposes. The bands used by the short-wave radiation scheme in the GCM are indicated by the short gray vertical lines.}
\label{fig:star_spectra}
\end{figure}

\subsubsection{Our Configurations}

In our experiments, the radius and the surface gravity of the planet are set to Earth's value. 

We only consider aquaplanets, i.e., planets whose surfaces are wholly covered with water. 
We use a thermodynamic ocean with 50 m depth for the fiducial runs for simplicity. 
However, in Section \ref{ss:sensitivity_ocean}, we also present the results of runs with a fully dynamic ocean, to show that our conclusions are not very sensitive to ocean heat transport. 

Our atmosphere has a mean surface pressure of 1 bar, and is composed mainly of N$_2$, with only 1 ppm CO$_2$, as in \citet{Kopparapu2016}. 
In Section \ref{ss:sensitivity_CO2}, we will briefly discuss the dependence of our results on the mixing ratio of CO$_2$. 
We did not include the effects of aerosols and photochemistry. 

The model was configured to have a horizontal grid with 4 degree resolution in latitude and 5 degree resolution in longitude, and 40 vertical layers covering up to 0.14 mbar (corresponding to $\sim 65$ km in the case of the Earth). 

In our experiments, we mainly changed two parameters: the total incident flux and the spectral type of the star. 
In the following, we represent the incident flux normalized by the solar constant by $S_X$.
We consider four types of stars: 
Sun (G2V), 
HD 22049 (K2V), 
Kepler-186 (M1V), and 
GJ~876 (M4V). 
The spectra of HD~22049 and GJ~876 are obtained from \citet{Segura2003} and \citet{Domagal-Goldman2014}, respectively, 
while the spectrum of Kepler-186 is a modeled spectrum from BT-Settl \citep{Allard2012} based on the stellar parameters reported in Table S1 of \citet{Quintana2014}. 
The physical parameters of these stars, as well as the references, are summarized in Table \ref{tbl:stellar_properties}, and the spectra are displayed in Figure~\ref{fig:star_spectra}. 

Throughout the paper, we exclusively consider synchronously rotating planets due to their relevance to future transmission observations of close-in systems, as discussed in Section~\ref{s:intro}. 
The orbital period (identical to the spin period) is changed according to Kepler's 3rd law, to be consistent with the total incident flux the planet receives and the stellar mass. 
Thus, the incident flux and the spin period are changed simultaneously for the fiducial set of runs. 
The orbital/spin periods corresponding to $S_X=1$ are shown in Table~\ref{tbl:orbital_period}. 
We also isolate the effect of the spin period in Section \ref{ss:sensitivity_Porbit}. 

Our fiducial models with a thermodynamic ocean typically equilibrate in $\sim $ 30 Earth years in model time. 
After the steady-state is reached, we average the 3D atmospheric profiles over an integer number of orbits that is approximately equal to 10 Earth years. 
All of the results presented in this paper are based on the averaged data created this way.

\subsubsection{Cloud Scheme}

In the ROCKE-3D GCM, Convection is triggered when air lifted from one model layer to the next higher layer becomes buoyant, according to a definition that includes both temperature and the molecular weight of H$_2$O relative to dry air in assessing whether the lifted air is less dense than the environment. 
The cumulus parameterization uses a mass-flux closure determined by the mass required to establish neutral buoyancy at a cloud base over a specified adjustment time. 
The total mass flux is divided into two components (``plumes'') that entrain at different rates \citep{DelGenio2007}. 
The plume rises until the updraft speed decreases to zero. 
Entrainment is proportional to parcel buoyancy and inversely proportional to the square of updraft speed, following \citet{Gregory2001}. 
Mass is detrained at all levels above the level of neutral buoyancy. 
A downdraft forms at any level when an equal mixture of cloud and environment is negatively buoyant. 
Updraft convective condensate is partitioned into fractions that precipitate, detrain, and are carried upward by assuming a Marshall-Palmer particle size distribution and empirical size-fall speed relationships and calculating the fraction of the size distribution whose fall speed exceeds, is comparable to, or is less than the diagnosed updraft speed \citep{DelGenio2005}. 

Detrained condensate is treated by the stratiform cloud microphysics \citep[an updated version of][]{DelGenio1996}. 
Stratiform cloud fraction is diagnosed as a function of relative humidity and stability, while condensed water evolves using a prognostic equation that includes representation of all relevant microphysical source and sink processes.  
Partitioning between the liquid and ice phase is specified as a function of temperature in the mixed phase temperature range, and then modified by a parameterization of diffusional growth of ice when snow falls into a supercooled liquid water layer.

\subsubsection{Radiation Scheme}
\label{sss:radiation}

For the radiation scheme, we adopt the Suite of Community RAdiative Transfer codes based on Edwards and Slingo\footnote{\url{https://code.metoffice.gov.uk/trac/socrates}}~\citep[SOCRATES,][]{Edwards1996,EdwardsSlingo1996} as described in \citet{Way2017}.
This radiation scheme uses the two-stream approximation of \citet{Zdunkowski1985} with a diffusivity of $D=1.66$ for the long-wave (thermal) component, and of \citet{Zdunkowski1980} with $D=2$ for the short-wave (stellar) component. 
Opacities are treated using the correlated-$k$ method \citep{Goody1989,Lacis1991}, with $k$-terms derived using exponential sum fitting of transmissions~\citep{Wiscombe1977} based on HITRAN2012~\citep{Rothman2013}. 
H$_2$O continuum absorption is treated using the CAVIAR continuum \citep{Ptashnik2011}. 
Both H$_2$O and CO$_2$ lines are cut off at 25 cm$^{-1}$, and the pedestal is removed from the H$_2$O lines as required by the water vapor continuum parameterization.
Overlapping gaseous absorption is treated using equivalent extinction \citep{Edwards1996,Amundsen2016}. 

The optical properties of liquid clouds were derived using Mie theory, with the effective radius diagnosed from the cloud water content assuming a cloud droplet number concentration of 60 cm$^{-3}$ for liquid clouds. 
The optical properties of ice crystals are described in \citet{Edwards2007}, and are based on the representation of ice aggregates introduced by \citet{Baran2001}, with the effective dimension calculated using a parameterization based on temperature. 
Convective and stratiform clouds are treated separately, and if both are present in the same grid box they are treated as randomly overlapping to calculate the total cloud fraction within the grid box. 
Vertical cloud overlap is treated using the maximum-random overlap assumption, where clouds in adjacent layers overlap maximally, while clouds separated by one or more layers overlap randomly. 

The configuration of the long-wave radiation scheme adopted here is the same as that used by the UK Met Office for global atmosphere configuration 7.0 (GA7.0; D. Walters et al. 2017, in preparation).  
For short-wave radiation, we use 29 bands, as indicated by the gray vertical lines in Figure ~\ref{fig:star_spectra} and listed in Table \ref{tab:short-wave}, in order to ensure accuracy of the short-wave absorption. 
The wavelengths are weighted internally in each band by each stellar spectrum when deriving $k$-coefficients and cloud optical properties to further improve accuracy. 
We check the accuracy of the short-wave radiation scheme by comparing the flux and heating rate obtained with $29$ bands to higher-resolution calculations, and find a good match; see the Appendix \ref{ap:radiation} for more details. 

\begin{figure*}[t]
    \begin{minipage}{0.5\hsize}
\includegraphics[width=\hsize]{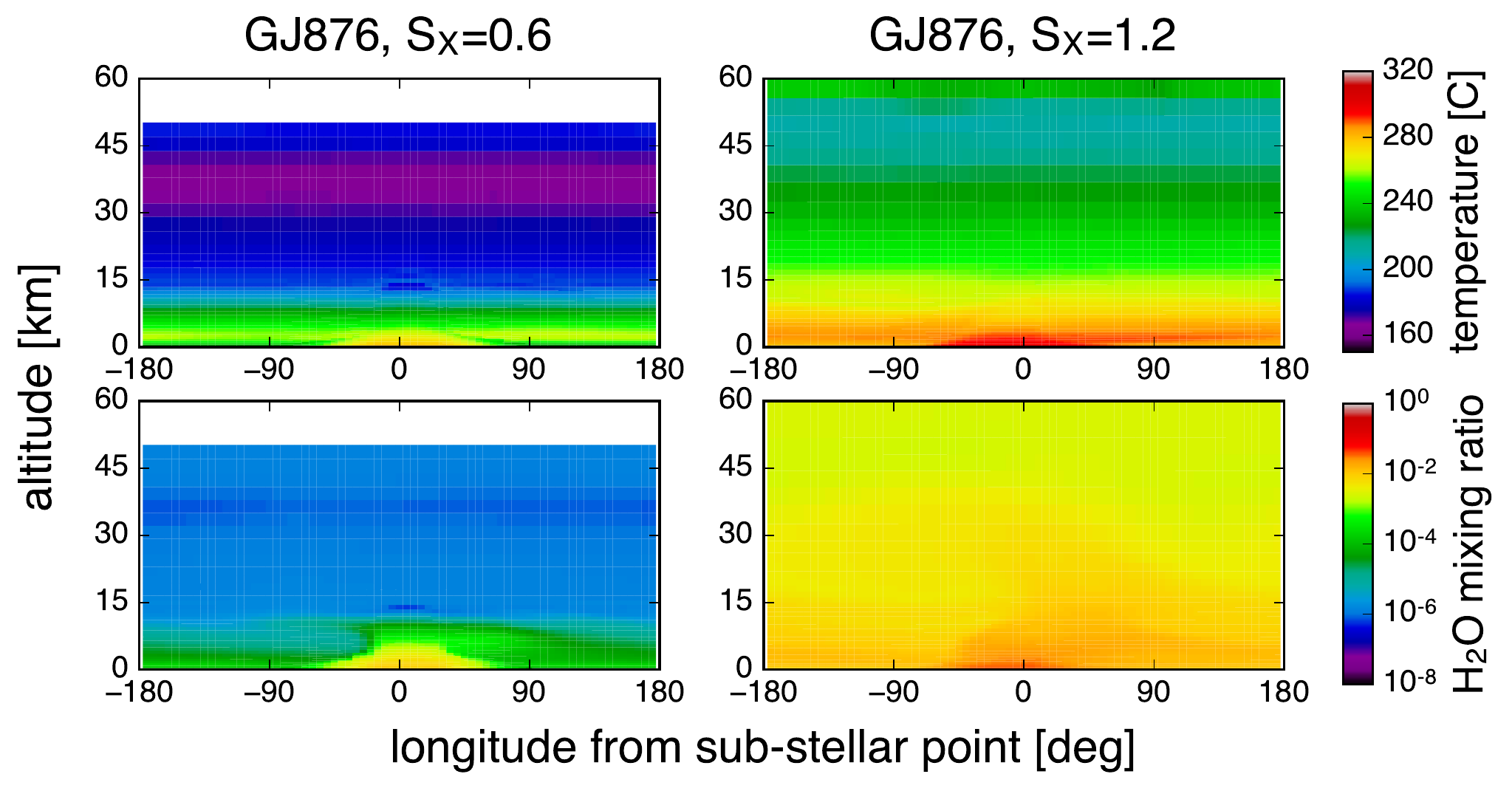}
    \end{minipage}
    \begin{minipage}{0.5\hsize}
\includegraphics[width=\hsize]{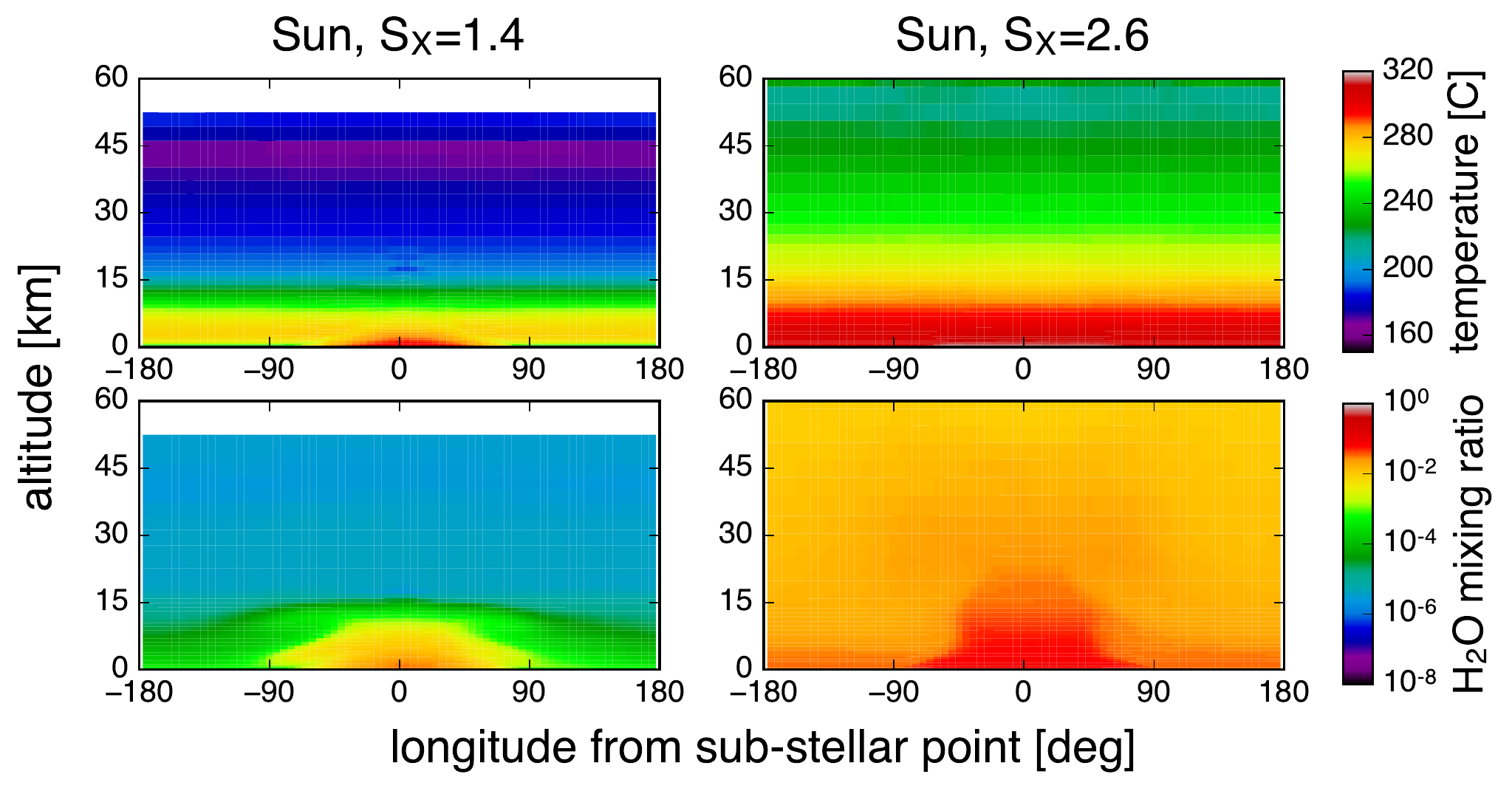}
    \end{minipage}
    \caption{Profiles of temperature (upper panels) and \wv{} mixing ratio (lower panels) along the equator, for GJ~876 (a M4V star; left) and for the Sun (right), with different total normalized incident flux ($S_X$). The blank area near the top is above the model top ($< 0.14$ mbar).}
\label{fig:3Dprofile_equator}
\end{figure*}

\begin{figure*}[t]
    \begin{center}
    \includegraphics[width=\hsize]{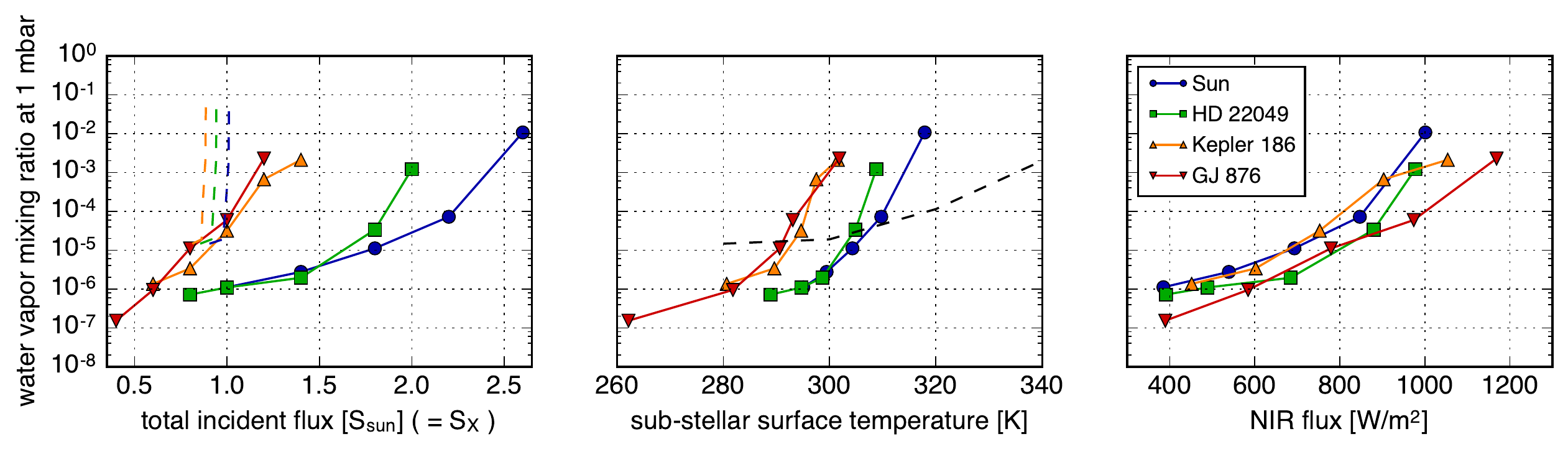}
    \end{center}
\caption{Water vapor mixing ratio at \preslevel{} (\xwater{}) as a function of total incident flux (left), the surface temperature averaged within $0.2R_\earth$ of the substellar point (center), and the near-infrared portion of the incident flux integrated between 0.9 and 3.0 \textmu m (right). The colors represent the stellar types: Sun (navy), HD~22048 (green), Kepler-186 (orange), and GJ~876 (red). The dashed lines in the left panel are adapted from the 1D model of \citet{Kopparapu2013} for varying stellar effective temperatures: 5800~K (navy), 4800~K (green), and 3800~K (orange). The dashed black line in the middle panel is also adapted from \citet{Kopparapu2013}. }           
\label{fig:xH2O_S0X}
\end{figure*}

\subsection{Simulation of transmission spectra}
\label{ss:method_TransmissionSpectra}

We also simulate transmission spectra based on the outputs of our GCM experiments, setting the modeled planet in a transiting geometry, in order to quantify the absorption features of \wv{}.  
We follow the procedure described in Appendix B of \citet{Way2017}, which is briefly summarized below. 

First, we obtain the vertical profiles of temperature, water vapor, and cloud particles along the terminator (i.e., dayside-nightside boundary) by linearly interpolating the profiles of the GCM gridpoints adjacent to the terminator. 
Then, we trace the optical paths of the transmitted ray according to the profile of the refraction index, from the observer toward the stellar disk, in the same way as described in \citet{vanderWerf2008}, following \citet{Misra2014}. 
After that, the spectral opacity at different points along the path is calculated considering Rayleigh scattering, the absorption by atmospheric molecules, and absorption and scattering by cloud particles, excluding the effects of multiple scattering. 
The spectral signatures of the planetary atmosphere are discussed in terms of the transit depth $\Delta F(\lambda )$ where $\lambda $ is the wavelength, as well as the effective height, $h_{\rm eff}$, which is given as a function of planet radius, $R_{\rm p}$, and stellar radius, $R_{\star }$, by 
\begin{equation}
\Delta F(\lambda) = \frac{( R_{\rm p} + h_{\rm eff}(\lambda)  )^2}{R_{\star }^2}. 
\end{equation}
To compute the absorption coefficients of atmospheric molecules, we use line-by-line opacity tables based on the HITRAN2012 database \citep{Rothman2013}, prepared separately from the opacities used in the GCM. 

Including the optical effects of clouds is tricky. 
Strictly speaking, one needs to know the size distribution and the spatial distribution of the cloud particles at the time scale of the planetary transit. 
In this paper, when calculating transmission spectra, we consider the effect of clouds only in an approximate manner with four assumptions. 
First, we assume that the radius of the cloud particles $r_{\rm cld}$ is $r_{\rm cld}=30~{\rm \mu m}$ (regardless of their phase), similar to the typical values of ice cloud particles on the Earth as well as in our GCM simulations.  
Second, we assume that the scattering cross section is given by that of the geometric optics limit, i.e. $2 \pi r_{\rm cld}^2$. 
Third, we assume that cloud particles are uniformly distributed within each grid cell. 
Fourthly, we compute the transmitted light based on the optical properties averaged over 10 Earth years, rather than compute it based on the instantaneous optical properties before taking the temporal average. 
The third and fourth points above imply that the effects of clouds we find are the largest among the possible realizations given the particle size distribution, because of the nonlinear dependence of transmittance on the optical thickness. 
Under these assumptions, the absorption coefficient $k$ of the clouds is 
\begin{equation}
k = 2 \pi r_{\rm cld}^2 \cdot n = 3 \rho x_{\rm cld} / ( 2 \rho_{\rm H2O} r_{\rm cld} ),
\end{equation}
where $n$ is the number density of cloud particles, $\rho $ is the density of the ambient atmosphere, $x_{\rm cld}$ is the cloud water content provided by the GCM experiments, and $\rho_{\rm H2O}$ is the density of liquid/solid water, which is assumed to be $1 {\rm g/cm}^3$. 
The uncertainties in the particle radius, scattering cross section, and the cloud water content could change the absorption coefficient by a factor of a few.

\section{Results}
\label{s:results}

\subsection{General Trend in 3D Atmospheric Structures}
\label{ss:result_H2Omixingratio}

Before examining the dependence of the \wv{} mixing ratio in the upper atmosphere, we discuss the 3D structures of temperature and water vapor, as we need to know at which altitude the mixing ratio of \wv{} becomes relatively uniform and hence
can be regarded as representative of the humidity in the upper atmosphere. 

Figure~\ref{fig:3Dprofile_equator} shows the structures of temperature and  \wv{} mixing ratio along the equator for the planets around GJ~876 (M4V) and around the Sun (G2V), respectively, for varying incident flux normalized by the solar constant ($S_X$). 
In all panels the substellar point corresponds to $0^{\circ }$ longitude. 
In general, the temperature and the \wv{} mixing ratio near the surface strongly depend on the distance from the substellar point, but are fairly uniform in the upper part of the atmosphere. 
Above about 1~mbar, which is located between 35 and 50 km altitude depending on the atmospheric temperature profiles, the \wv{} mixing ratio becomes horizontally uniform within a few tens of percent. 
Therefore, in the following, we denote the globally averaged \wv{} mixing ratio at \preslevel{} by \xwater{}, and use it as representative of the \wv{} abundance in the upper atmosphere. 

We also observe optically thick clouds around the substellar point and a resulting increase of the planetary albedo as a function of total incident flux (not shown in figures), consistent with \citet{Yang2013,Yang2014}, \citet{Kopparapu2016}, and \citet{Way2016}. 
As shown in Figure~\ref{fig:3Dprofile_equator}, the maximum surface temperature does not exceed 320~K even in the highly irradiated cases in our experiments where the \wv{} mixing ratio near the model top is as high as $10^{-3}$; we will discuss this in the subsequent sections. 

\begin{figure*}[tbh]
    \begin{center}
    	\includegraphics[width=0.9\hsize]{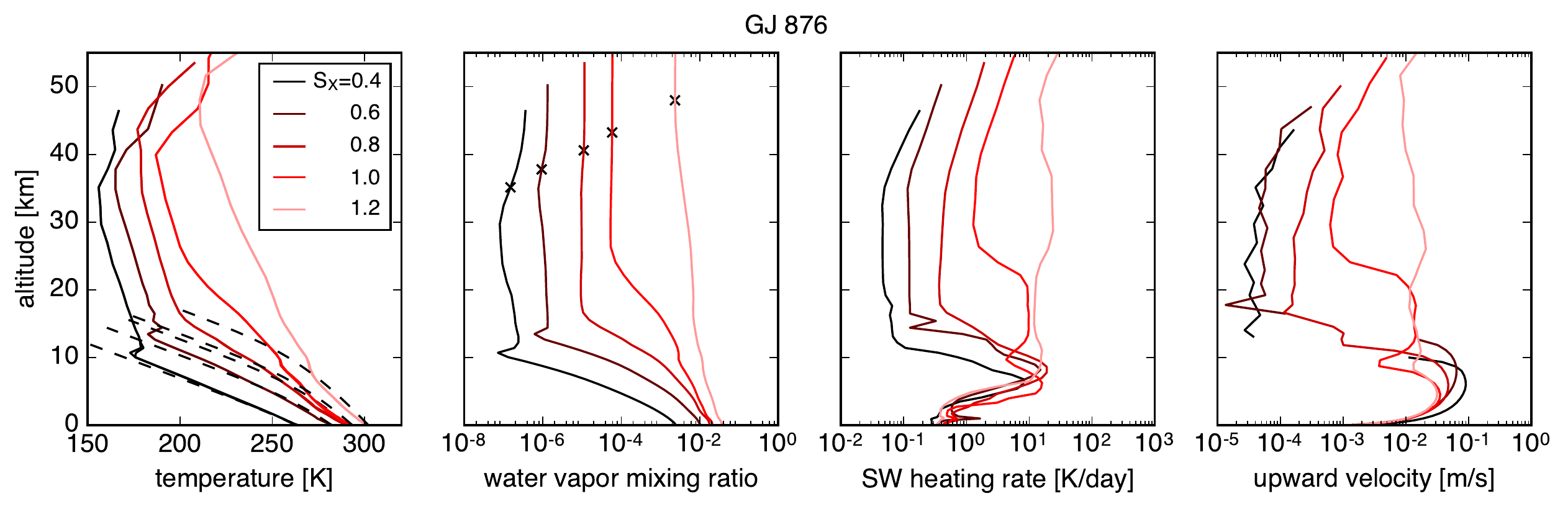}
    \end{center}
    \begin{center}
        \includegraphics[width=0.9\hsize]{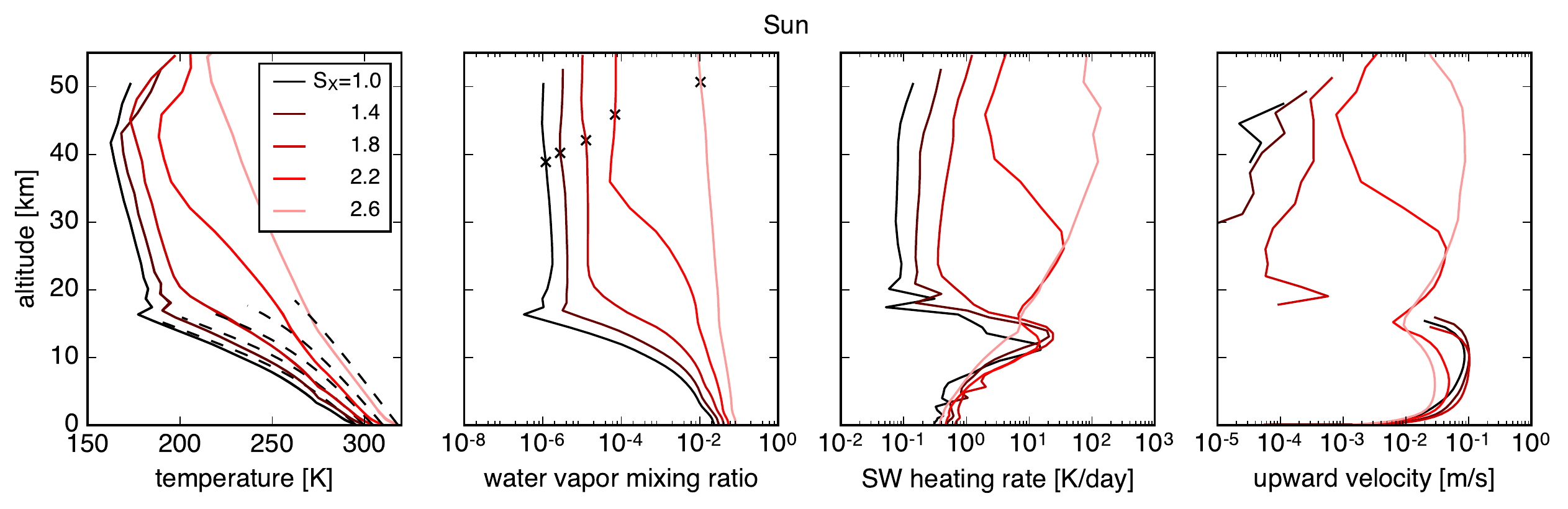}
    \end{center}
\caption{From left to right, the vertical profiles of temperature, \wv{} mixing ratio, short-wave heating rate, and upward velocity, averaged within $0.2R_\earth$ of the substellar point, for a planet around GJ~876 (top) and around the Sun (bottom), with varying total incident flux. The dashed lines in the leftmost panels show the moist adiabat drawn from the same surface temperature. The cross marks indicate the altitude of \preslevel{} where \xwater{} is measured. In the rightmost panel, the altitudes with negative upward velocity are not shown. }
\label{fig:AqOH0TLS_GJ876_temp_xH2O_vz_heat}
\end{figure*}

\subsection{Response of Water Vapor Mixing Ratio \\to Increased Incident Flux}
\label{ss:result_H2Omixingratio}

In this section, we examine in more detail how the \wv{} mixing ratio in the upper atmosphere depends on the incident flux (both the total flux and the spectral distribution). 

The left panel of Figure~\ref{fig:xH2O_S0X} presents \xwater{} versus $S_X$. 
Overall, we find an increase of \xwater{} as a function of $S_X$. 
However, the slope is much more gradual than suggested by 1D models, in particular those  of \citet{Kasting1993} and \citet{Wordsworth2013} with a small CO$_2$ mixing ratio, which typically predict a rapid transition from an Earth-like dry stratosphere to an \water{}-dominated atmosphere. 
Our result that the \wv{} mixing ratio responds more slowly to the incident flux is qualitatively consistent with \citet{Yang2013}, who studied synchronously rotating planets using CAM3 with an emphasis on the stability of the climate. 
Note that we cannot directly compare to their results, as our assumptions do not match perfectly; specifically, \cite{Yang2013} assumed a planet with a larger radius ($R=2R_\earth$) and higher gravity ($g=1.4g_\earth$), and a 60-day orbital period. 
The dependence on these other planetary parameters will be investigated in a future paper. 

Because these are synchronously rotating planets where the surface temperature is stabilized by the optically thick clouds around the substellar point, one might think that this stabilized surface temperature would explain the slow  response of \wv{} to the increased total incident flux. 
To see the effect of the surface temperature, we have plotted \xwater{} as a function of surface temperature around the substellar point, where convection is strongest (Figure~\ref{fig:xH2O_S0X}, middle panel). 
It is then clearly seen that the \wv{}  mixing ratio starts to rise at a fairly modest surface temperature and increases even more dramatically as a function of surface temperature than 1D models suggest. 
This indicates that the increasing surface temperature cannot by itself explain the increased stratospheric \wv{} mixing ratios. %

Instead, we find that the near-infrared (NIR) component of the incident flux appears to be a good predictor of stratospheric humidity across the full range of stellar spectral types (Figure~\ref{fig:xH2O_S0X}, the right panel). %
Here, we integrated the star spectra between 0.9 and 3 \textmu m (the interval is somewhat arbitrary) to obtain the NIR component of the incident flux. %
We will examine the reason for this in the next subsection.

\subsection{Development of Large-scale Circulation \\in the Upper Atmosphere}
\label{ss:result_omega}

In Figure~\ref{fig:AqOH0TLS_GJ876_temp_xH2O_vz_heat}, we show the vertical profiles of temperature, \wv{} mixing ratio, short-wave heating rate, and upward velocity, averaged over the region within a distance $0.2R_\earth$ from the substellar point. 
The variability of the profiles in the individual gridboxes that contribute to the average is negligible compared to the changes due to varying $S_X$. 
Rising motion exists at most altitudes in the substellar region.  In the troposphere, where moist convection occurs, large-scale vertical motion is relatively insensitive to $S_X$.  Above the convective layers, though, large-scale upward motion increases strongly with incident flux, concurrently with the increase in \wv{} mixing ratio and heating rate in the upper atmosphere. 

Here in the upper atmosphere we see the interplay among the radiative heating, water abundance, and the vertical air transport:  
when the incident flux increases, the radiative heating due to the absorption of short-wave radiation by \wv{} and potential cloud particles, mainly in the NIR regime, also increases. %
The air then reacts by rising and adiabatically cooling. %
The induced upward motion of the atmosphere transports water vapor from the moist lower atmosphere to the upper atmosphere, lowering the vertical gradient of water vapor and increasing the moisture at high altitude, further increasing the short-wave heating. 

\begin{figure}[tb]
    \begin{center}
\includegraphics[width=\hsize]{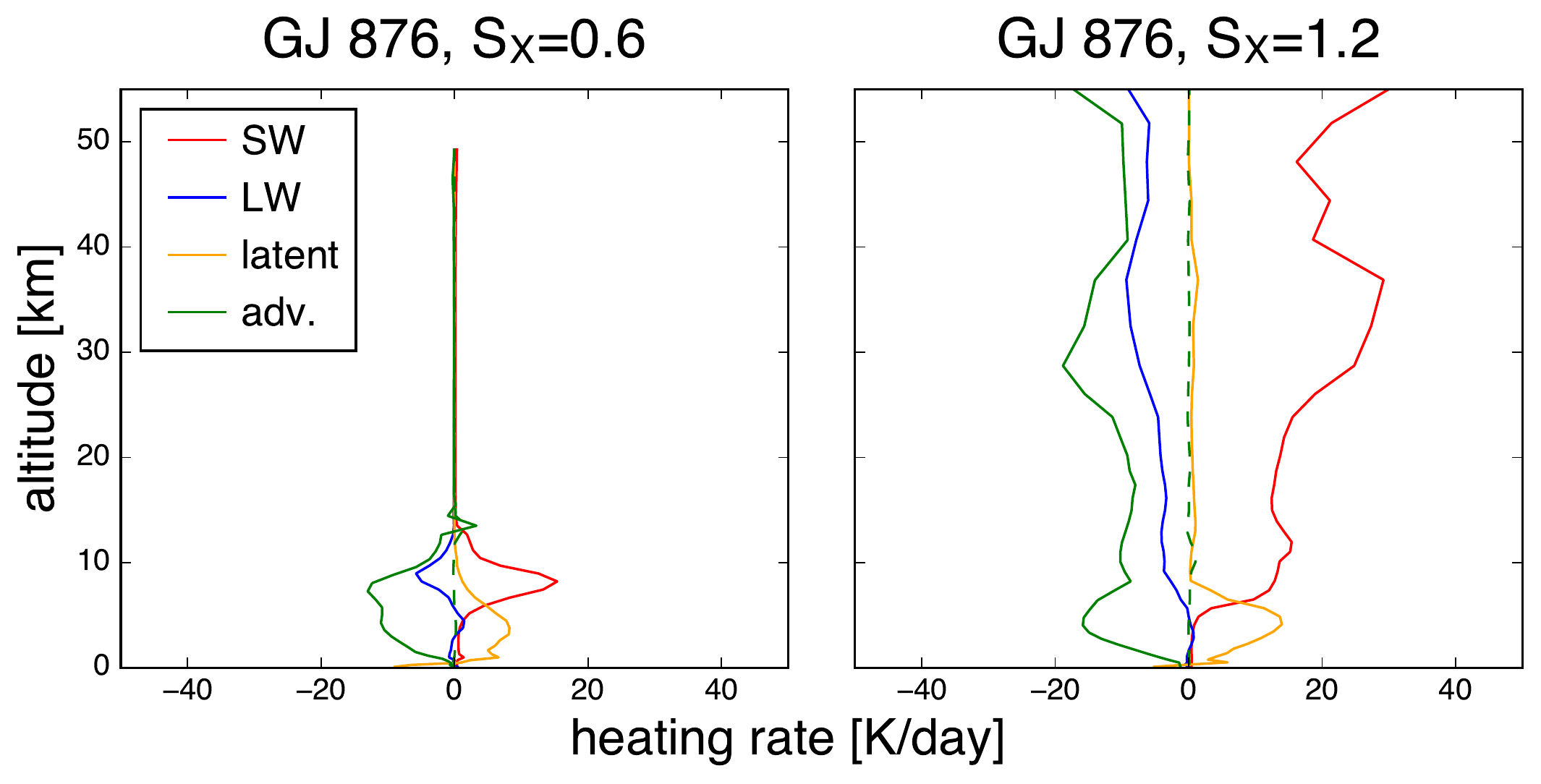}
\includegraphics[width=\hsize]{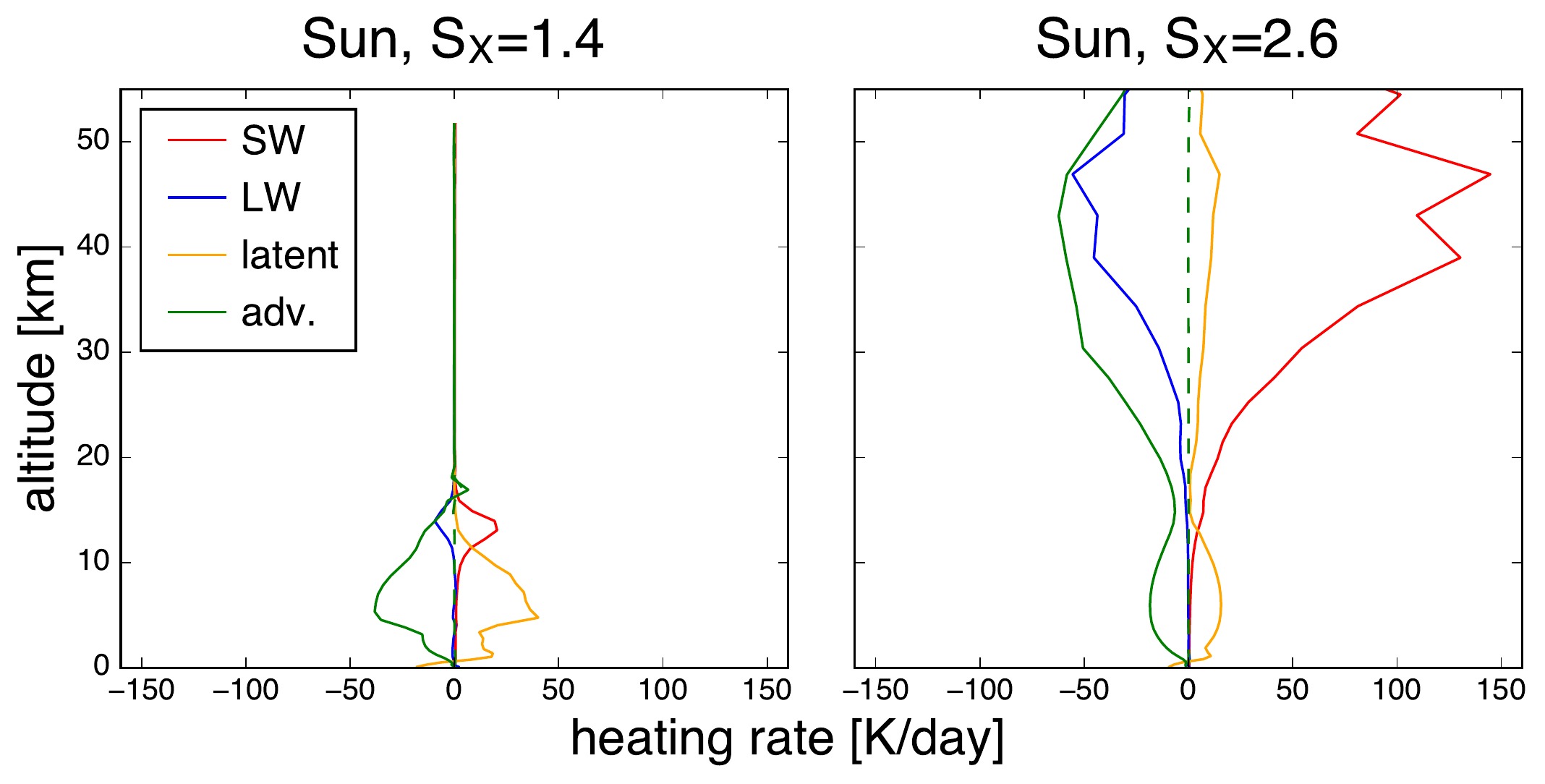}
    \end{center}
    \caption{Contribution of different physical processes to the heating rate in the substellar regions in the cases of low (left) and high (right) irradiance, for GJ~876 (top) and the Sun (bottom). Red: heating rate from short-wave radiation, blue: long-wave radiation; orange: latent heat; green solid: vertical advection; green dashed: horizontal advection.}
\label{fig:heating_rate}
\end{figure}

To be more specific, the evolution of the profile satisfies the conservation equation for the potential temperature: 
\begin{equation}
\frac{T}{\theta } \left( \frac{\partial \theta }{\partial t} + ( {\bf v} \cdot \nabla ) \, \theta \right) = \frac{ Q_r + Q_l }{\rho c_p} ,   \label{eq:conserv_energy}
\end{equation}
or
\begin{equation}
\frac{T}{\theta } \frac{\partial \theta }{\partial t} = - \frac{T}{\theta }\left( u \frac{\partial \theta }{\partial x} + v\frac{\partial \theta }{\partial y} \right) -\frac{T}{\theta } \left(  \omega \frac{\partial \theta }{\partial p} \right)  + \frac{ Q_r }{\rho c_p} + \frac{ Q_l }{\rho c_p}, \label{eq:conserv_energy_2}
\end{equation}
where $p$ is the pressure, $Q_r$ and $Q_l$ are the radiative and latent heating rates, $c_p$ is the specific heat of the atmosphere at constant pressure, $\rho $ is atmospheric density, $\{ u, v, \omega \}$ are the conventional velocity components toward the east ($x-$axis), north ($y-$axis), and upward ($z-$axis; expressed here in pressure coordinates as $\omega \equiv dp/dt$), respectively. 
Given that the horizontal advection (the first term on the right side) has a negligible contribution since the horizontal gradients are small (Figure~\ref{fig:3Dprofile_equator}), and that the latent heating (the last term on the right side) is also negligible above the altitude to which convection penetrates, the increase of the radiative heating in the upper atmosphere must be balanced by the adiabatic cooling by vertical advection. %
The contribution of each term of the right side of equation (\ref{eq:conserv_energy_2}) in the steady state is shown in Figure \ref{fig:heating_rate}, for the substellar regions of low and high-irradiation cases with GJ~876 and the Sun. %
It is clearly seen that the net radiative heating in high-irradiation cases is compensated by the cooling by the vertical advection. 

\begin{figure}[htb]
    \begin{center}
    \includegraphics[width=1\hsize]{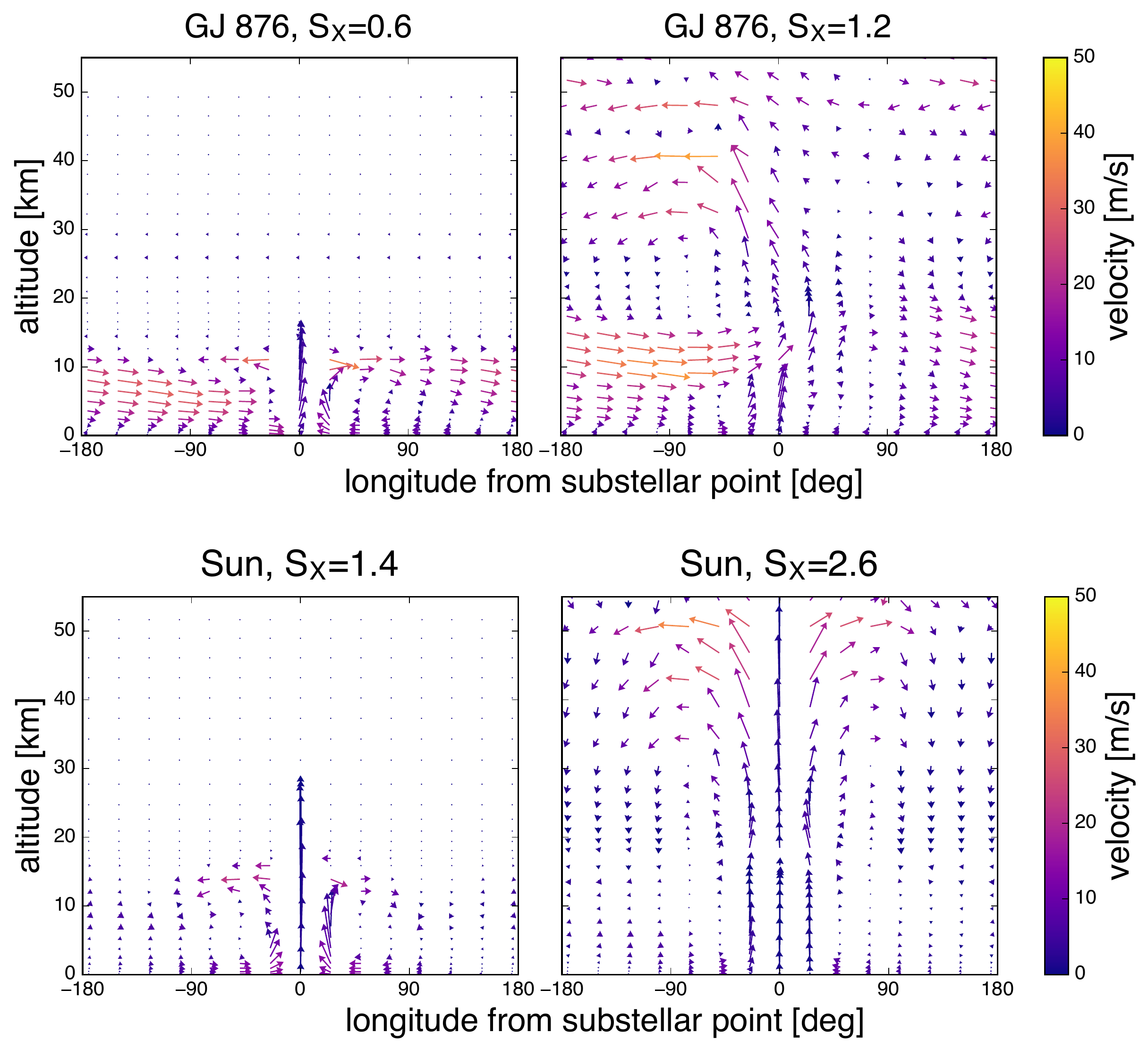}
    \end{center}
\caption{Velocity fields along the equator in the cases of low (left) and high (right)irradiance, for GJ876 (top) and the Sun (bottom). The vertical components of the vectors are multiplied by 1000 for display purposes, while the colors show the absolute values, $\sqrt{v_x^2 + v_z^2} \sim  v_x$. }
\label{fig:vectormap}
\end{figure}

\begin{figure*}[htb]
    \begin{center}
    \includegraphics[width=1\hsize]{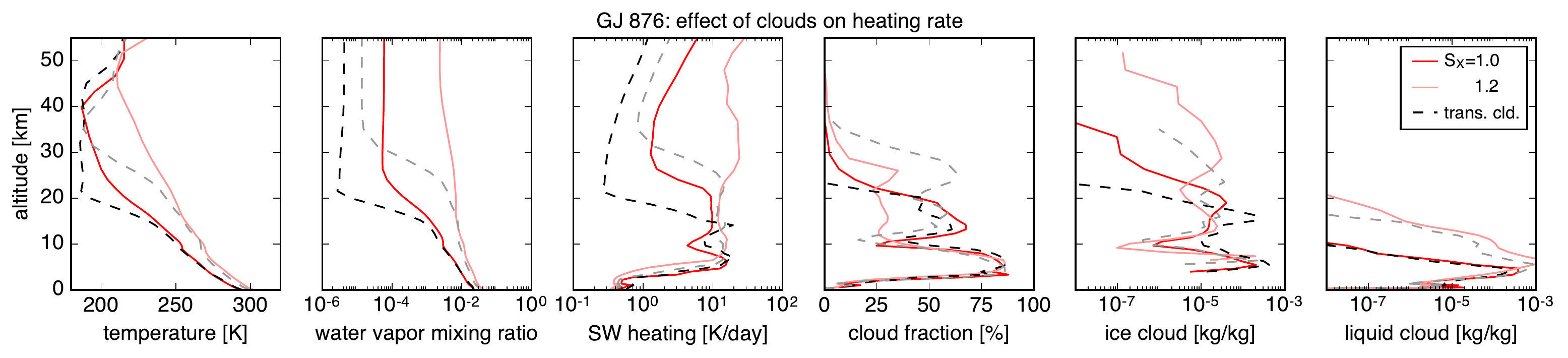}
    \end{center}
\caption{Comparison between the fiducial $S_X=1.0$ (red) and $S_X=1.2$ (pink) models of of planets orbiting GJ~876 and the comparable ``transparent high clouds'' models where the radiative effect of high clouds above 237 mbar are turned off (dashed black for $S_X=1.0$ and dashed gray for $S_X=1.2$). The vertical profiles of temperature, \wv{} mixing ratio, short-wave heating rate, cloud fraction, ice cloud content, and liquid cloud content are averaged over the region within $0.2R_\earth $ of the substellar point. }
\label{fig:GJ876_heat_cld}
\end{figure*}

Figure \ref{fig:vectormap} puts this upward motion in the substellar regions into a global context by presenting the velocity fields in the altitude-longitude plane along the equator for the low and high-irradiation cases for GJ~876 and the Sun. %
Here, the vertical components of the vectors are multiplied by 1000 to make them more visible, because large-scale vertical motions are about three orders of magnitude weaker than horizontal motions. %
The development of the large-scale circulation above the troposphere is clearly seen. 

Altogether, Figures \ref{fig:AqOH0TLS_GJ876_temp_xH2O_vz_heat}, \ref{fig:heating_rate} and \ref{fig:vectormap} suggest that the stratospheric dynamical response to increased stellar radiation is similar for stars of different spectral type, but for a warmer star the onset of strong upward motion occurs at a much larger incident flux because less of the instellation is in the NIR where the radiation is more efficiently absorbed by \wv{} and clouds. %
This is the reason for the behavior seen in the right panel of Figure~\ref{fig:xH2O_S0X}. 

We also note that the contribution of radiative heating by cloud particles above convective layers is not negligible. 
In order to see the radiative effect of clouds above convective layers, we created models that are comparable to the fiducial models, but where the  influence of cloud particles above the 237 mbar level ($\sim 11$ km for the runs shown below) has been turned off in the radiation calculations (i.e., cloud particles in the upper atmosphere are transparent to radiation). 
Figure~\ref{fig:GJ876_heat_cld} displays the comparison between such ``transparent high clouds'' models (dashed black, dashed gray) and fiducial models (red, pink), for $S_X=1.0$ and $1.2$ with the incident spectrum of GJ~876. 
While these high clouds have a minimal effect on the surface temperature and atmospheric profiles in the convection layers, it is seen that radiative heating above the convection layer is substantially suppressed without the radiative effect of clouds, leading to a colder, and drier upper atmosphere. 
Note that we still see heating attributed to water vapor in the ``transparent high clouds'' models; for example, the \wv{} mixing ratio and short-wave heating at 20-30 km altitude increase considerably from $S_X$ = 1.0 to $S_X$ = 1.2, even in the absence of cloud radiative effects, but the sensitivity to $S_X$ is weaker at higher altitudes. 
We also ran equivalent simulations for other types of stars and found that the overall effects of the ``transparent high clouds'' assumption are very similar. 

In summary, atmospheric vertical motion and the radiative heating by \wv{} and cloud particles interact with each other through positive feedbacks, and thus a modest increase of the incident radiative flux, $S_X$, can result in  an order-of-magnitude or more increase of the \wv{} mixing ratio in the upper atmosphere, while the surface temperature increases only modestly due to the stabilization by clouds. 
This picture is verified by a good correlation between the \wv{} mixing ratio at \preslevel{} and the NIR portion of the incident flux as shown in the right panel of Figure~\ref{fig:xH2O_S0X}, insensitive to the stellar spectral types. %
Because the NIR flux is an external parameter that can be found from direct observations of the host star, it would be a useful guide to predict the high-altitude \wv{}  mixing ratio for a planet with surface liquid water. 

\subsection{Transmission Spectra}
\label{ss:result_TransmissionSpectra}

\begin{figure}[!b]
    \begin{center}
    \includegraphics[width=\hsize]{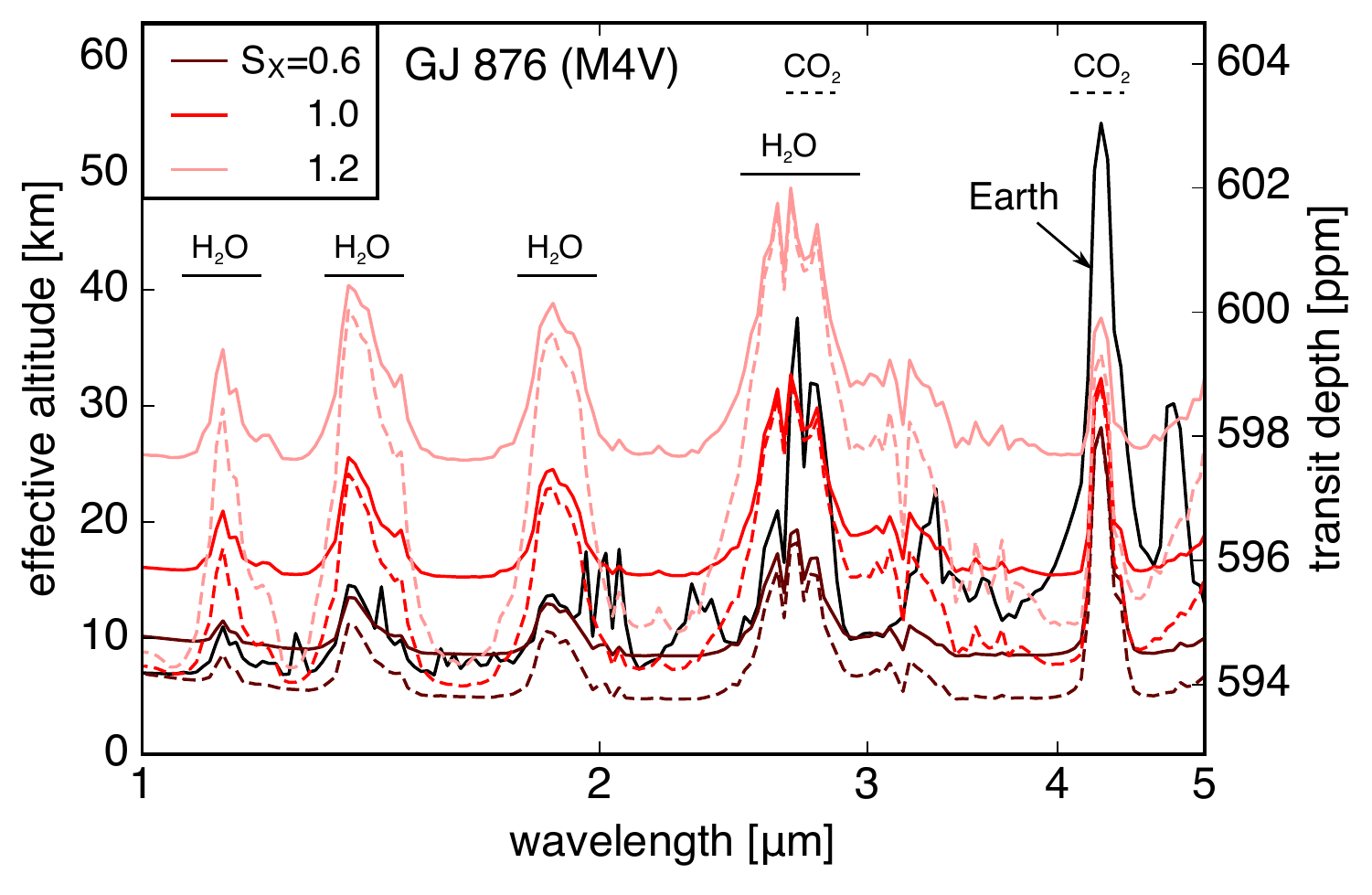}
    \end{center}
\caption{Modeled transmission spectra based on GCM experiments for planets around GJ~876 (M4V) with $S_X=0.6$ (red), $1.0$ (red) and $1.2$ (pink), with (solid) and without (dashed) clouds. The corresponding vertical profiles in the substellar region are shown in Figure~\ref{fig:AqOH0TLS_GJ876_temp_xH2O_vz_heat}. The left side of the $y$-axis shows the effective altitude, while the right side indicates the transit depth in ppm using the size of the star from Table~\ref{tbl:stellar_properties}. The spectral resolution is lowered to $\mathcal{R}=100$.  For a reference, the black solid line shows the modeled transmission spectrum of the Earth based on the US standard atmosphere. }
\label{fig:transmission}
\end{figure}

In this subsection, we show the dependence of the transmission spectra on the incident flux, taking the modeled planets around one of our M-type stars, GJ~876 as examples. 
Figure~\ref{fig:transmission} shows the transmission spectra for the $S_X=0.6$, $1.0$ and $1.2$ cases, where the \wv{} mixing ratio at \preslevel{} is roughly $10^{-6}$, $10^{-4}$ and $3\cdot 10^{-3}$, respectively. 
The corresponding vertical profiles in the substellar region are shown in Figure~\ref{fig:AqOH0TLS_GJ876_temp_xH2O_vz_heat}, and note that the \wv{} mixing ratio is even higher at medium altitudes (10-30~km).
Although refraction due to the planetary atmosphere is included, its effect on the transmission spectra of planets around late-type stars like GJ~876 is minor \citep{Betremieux2014,Misra2014}. 

Without the opacity due to clouds (colored dashed lines in Figure~\ref{fig:transmission}), the effective altitude of the water vapor absorption features increases from $\sim $15~km to $\sim $40~km when $S_X$ increases from $0.6$ to $1.2$. 
This level of effective altitude of water signatures is comparable to the strongest features (CO$_2$) in the modeled visible-to-NIR transmission spectrum of the Earth (solid black line) , and substantially larger than the H$_2$O signatures of the Earth. %
Note that the feature around 2.7\textmu m is the overlap between H$_2$O and CO$_2$, and in our simulation cases with 1 ppm of CO$_2$ the feature predominantly comes from H$_2$O, while in the case of the Earth it mostly comes from CO$_2$. 
Thus, for the highly irradiated cases shown here, \water{} signatures will be a natural target when trying to observe the strongest atmospheric signatures of planets similar to the Earth.

However, we also find clouds in the upper atmosphere around the terminator, and the inclusion of clouds substantially raises the baseline of the transmission spectra. 
(Note that the high clouds discussed in the previous sections are those around the substellar point, while what matter here for the transmission spectra are those at the terminator.) 
These clouds at the terminator presumably form from water vapor that is transported from the substellar point and cools radiatively or adiabatically. 
Although these tenuous clouds at high altitude would be optically thin vertically, the grazing geometry of transmission observations cause them to significantly affect the transmission spectra due to the long optical path length. 
After including the effect of clouds, however, we still find the overall increase of water vapor signatures as the incident flux increases. 
Compared to the standard Earth model, or the low-irradiation case (\wv{} mixing ratio $\sim 10^{-6}$) with the same assumptions for the cloud optical properties, the signal of a high-irradiation case could be larger by a factor of a few. 

We can evaluate the detectability of these features for idealized cases where the only source of observational uncertainty is photon noise. 
For an Earth-sized planet 10 pc away, assuming a 6.5~m telescope (having the James Webb Space Telescope in mind) with 40 \% throughput and a wavelength resolution $\mathcal{R}=$20, absorption features with 20~km differential effective altitude in this wavelength range (e.g., the feature around 2.7~\textmu m for the cases with clouds) would become detectable (5$\sigma $) after $\sim 100$ hr of integration with an M4 star (assuming blackbody radiation with $R_{\star }=0.4 R_{\odot }$, T=3500~K, similar to GJ~876), or after $\sim 15$ hr with an M8 star (with $R_{\star }=0.1 R_{\odot }$, T=2500~K, similar to TRAPPIST-1), respectively.

\section{Sensitivity Experiments }
\label{s:sensitivity}

In this section we examine the sensitivity of our results to some of our assumptions. 
In particular, we discuss the effect of the assumption of a simple thermodynamic ocean (Section \ref{ss:sensitivity_ocean}), the effect of the background CO$_2$ mixing ratio (Section \ref{ss:sensitivity_CO2}), the effect of the spin period (Section \ref{ss:sensitivity_Porbit}), and the effect of the uncertainty in the water vapor continuum absorption (Section \ref{ss:sensitivity_continuum}).

\subsection{Effect of Ocean Model}
\label{ss:sensitivity_ocean}

So far, we have used a thermodynamic ocean model with zero ocean heat transport and 50 m depth. 
Indeed, this simplified treatment has been regularly used in previous 3D atmospheric modeling studies of exoplanets. 
A more realistic approach is to include the dynamics of the ocean. 
Therefore, we check whether our results are sensitive to the prescription of the ocean by comparing our fiducial results and those obtained using a dynamic ocean module. 

For this purpose, we ran the models for GJ~876 assuming a 900 m deep dynamic ocean. 
The models with the dynamic ocean typically equilibrated after a hundred to several hundred Earth years. 
In the same way as the fiducial models, we averaged the atmospheric profiles  over approximately 10 Earth years after the model reached a steady state. 

\begin{figure}[!h]
    \begin{center}
    \includegraphics[width=0.9\hsize]{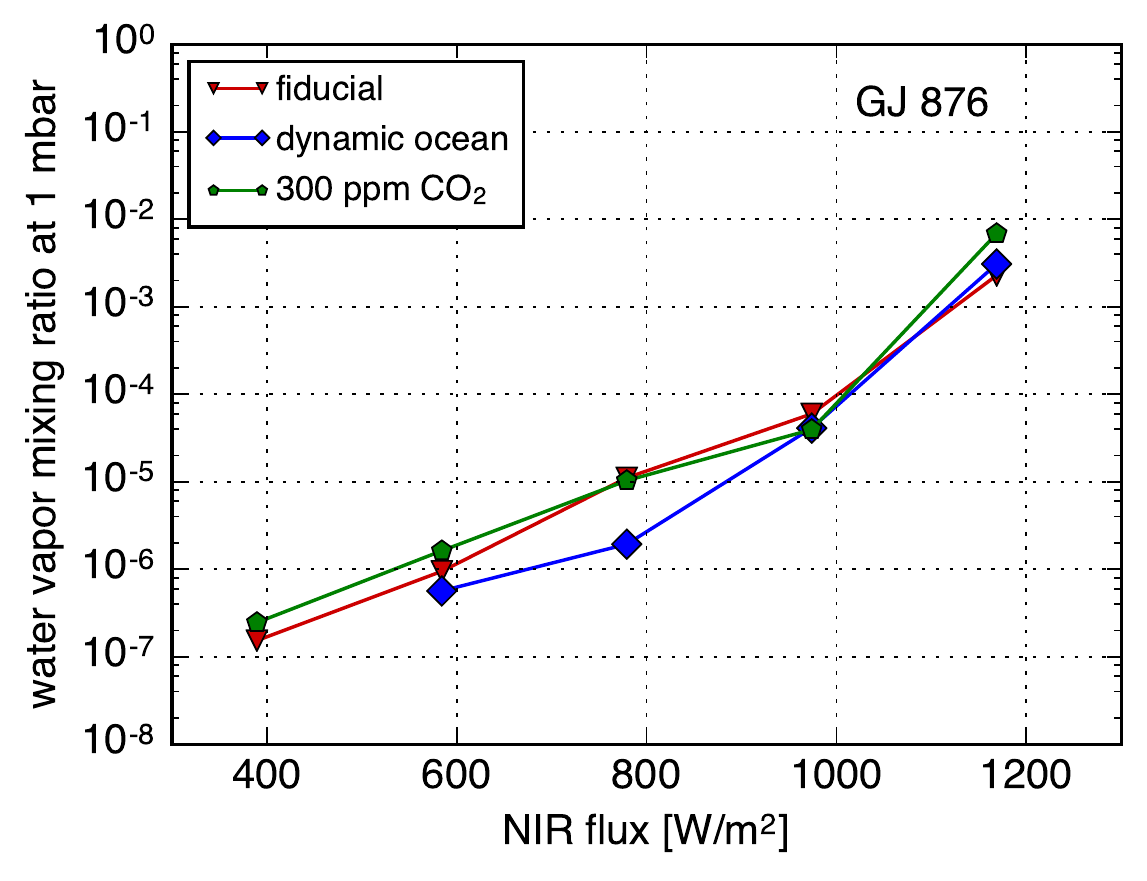}
    \end{center}
\caption{Water vapor mixing ratio at 1.39 mbar for the models using a dynamic ocean (blue diamonds) and the models with an Earth-like CO$_2$ mixing ratio (green pentagons), in comparison with the fiducial model with a thermodynamic ocean and 1 ppm CO$_2$ (red down-triangle). }
\label{fig:change_ocean}
\end{figure}

Figure~\ref{fig:change_ocean} shows \xwater{} for GJ~876 with varying $S_X$ both for our fiducial runs and those using the dynamic ocean module.  
There are minor differences between the two, with the maximum difference being less than an order of magnitude. 
The models with a dynamic ocean tend to have somewhat less \xwater{} than the models with zero ocean heat transport. 
This appears to be related to the fact that ocean heat transport from the dayside to the nightside lowers the substellar surface temperature and weakens convection there, and thus there is less transport of water vapor to the upper atmosphere. 

Overall, however, our fiducial models with a thermodynamic ocean give a reasonable order-of-magnitude estimate of \wv{} mixing ratio in the upper atmosphere. 

\subsection{Effect of CO$_2$}
\label{ss:sensitivity_CO2}

Our fiducial models assumed 1 ppm CO$_2$, which was an arbitrary choice. 
Although a full investigation of the dependence on the CO$_2$ mixing ratio is beyond the scope of this paper, here we briefly mention how much the results are changed if we assume an Earth-like CO$_2$ mixing ratio. 
The results of the models with 300 ppm of CO$_2$ are plotted in Figure~\ref{fig:change_ocean}. 
As shown, this level of variation of CO$_2$ has a very small effect on the \wv{} profile. 
Larger amounts of CO$_2$ up to that of a CO$_2$-dominated atmosphere could have a more substantial effect on the 3D profiles \citep{Wordsworth2013}.

\subsection{Effect of Spin Period}
\label{ss:sensitivity_Porbit}

In our fiducial set of experiments, we changed the planetary spin period together with the stellar flux for a given star, because we consider synchronously rotating planets and the orbital period (which is equal to the spin period) must change to be consistent with the total incident flux and the host star mass according to Kepler's 3rd law. 
Thus, strictly speaking, the change in the \wv{} mixing ratio presented in Figure~\ref{fig:xH2O_S0X} includes the effect of both the varying incident flux and the varying spin period. 
Meanwhile, it is well-known that the spin period has first order effects on the atmospheric circulation and hence can potentially affect \xwater{}. 

\begin{figure}[!t]
    \begin{center}
    \includegraphics[width=\hsize]{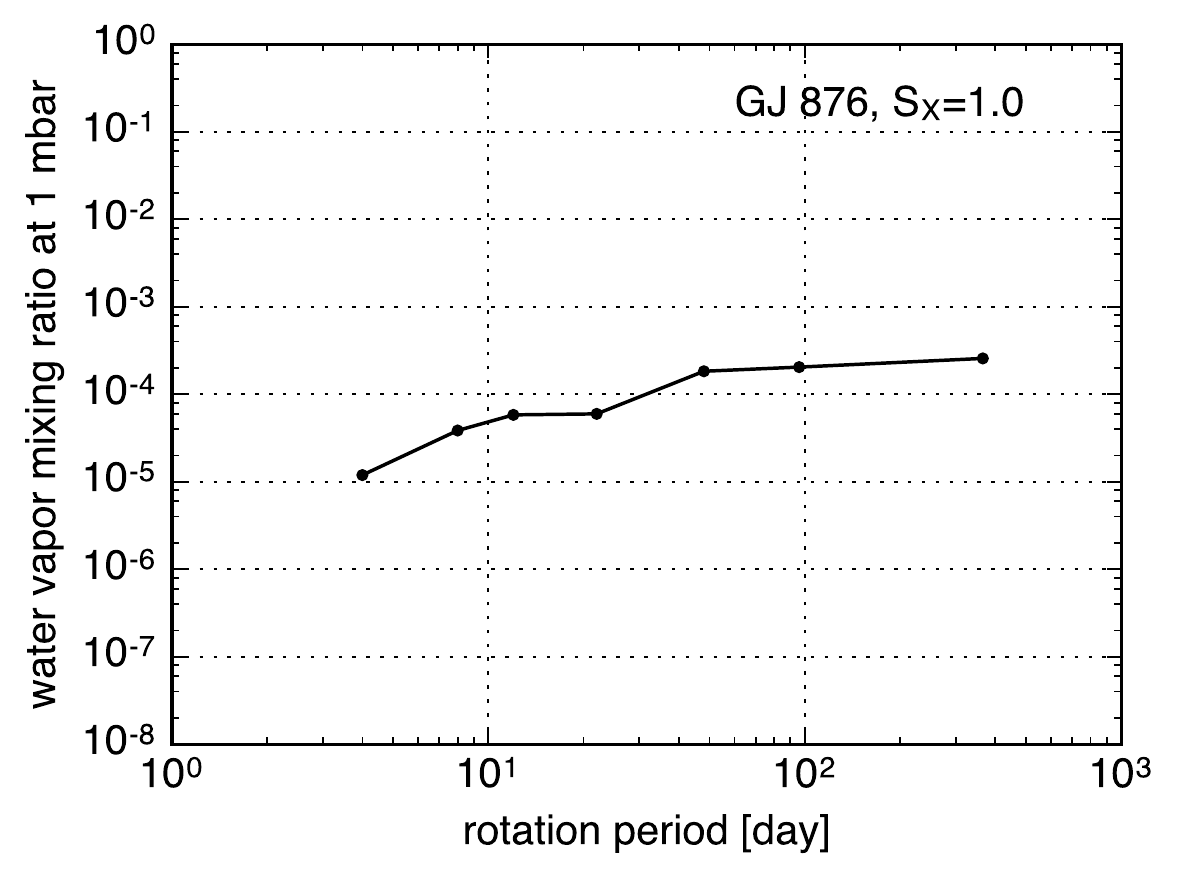}
    \end{center}
\caption{The dependence of \wv{} mixing ratio on the spin period. The host star spectrum is GJ~876 with $S_X=1$. In all cases, the planet is assumed to be synchronously rotating. }
\label{fig:changeP}
\end{figure}

In order to isolate the effects of rotation, we performed experiments where we varied the spin period while keeping the incident flux fixed at $S_X=1$ with the GJ~876 spectrum, but allowing the planet to remain in synchronous rotation. 
The effect of varying spin period on \xwater{} is presented in Figure~\ref{fig:changeP}. 
It is considerably smaller than the dependence on $S_X$ seen in Figure~\ref{fig:xH2O_S0X}, i.e., the increase in stratospheric water vapor as the planet is moved closer to its star is mostly an intrinsic sensitivity to the stellar flux rather than an effect of the changing planet rotation as the orbital period changes. 

The spin effect is non-negligible, however; we find that \xwater{} becomes smaller as the spin period becomes very short, i.e., increasing planet spin partly offsets the effect of increasing stellar flux. 
Our interpretation of this tendency is as follows. 
At slow rotation (long spin period) the day-night circulation dominates the atmosphere. 
As the spin period becomes shorter, however, the atmosphere develops high-altitude zonal winds, as demonstrated in previous papers \citep{Merlis2010, Edson2011, Kopparapu2016}. 
The super-rotating zonal wind seen in the equatorial region in the shorter period runs (not shown) efficiently transports water vapor from the dayside to the nightside, limiting the cloudiness in the substellar region, and resulting in a higher surface temperature \citep{Kopparapu2016}.  
Consistently, in the runs with a shorter spin period, the convection at the substellar point is deeper, and the temperature at the top of the convection  where it moistens the atmosphere is colder, and thus the convection is less able to moisten the atmosphere there. 
Moreover, the \wv{} transported upward by convection is more efficiently transported to the nightside by the zonal wind, as noted above; the water vapor mixing ratio on the nightside increases by about an order of magnitude when the rotation period decreases from 365 Earth days to 4 Earth days. 
These processes are likely to lead to the smaller upper atmosphere \wv{} mixing ratio observed in our shorter shorter period runs. 

\subsection{Water Vapor Continuum}
\label{ss:sensitivity_continuum}

We have shown that the radiative heating caused by absorption of the incident NIR radiation by water vapor is one of the key processes governing the upper atmosphere humidity, and the water vapor continuum absorption is partly responsible for this. 
There is significant uncertainty in the strength of both the self- and foreign-broadened water vapor continua at NIR wavelengths \citep{Campargue2016,Shine2016}. 
Our fiducial models used the CAVIAR water vapor continuum \citep{Ptashnik2011}, which from recent laboratory measurements appears to overestimate particularly the self-broadened water vapor continuum \citep{Shine2016}. 
Another popular continuum model, MT\_CKD \citep{Mlawer2012}, or its predecessor CKD \citep{Clough1989}, appears to be more in line with, or a bit below, the recent measurements of the self-broadened continuum. 
Measurements of the foreign-broadened continuum are sparse, but more consistent with CAVIAR as MT\_CKD (and CKD) appears to underestimate the strength of this continuum \citep{Shine2016}. 

In order to evaluate the impact of the water vapor continuum model on our results, we have also run simulations with the CKD 2.4 water vapor continuum. 
As CAVIAR and CKD 2.4 approximately bracket the available measurements, this should give a good indication of the sensitivity of our results to the adopted continuum model. 

\begin{figure}[!t]
    \begin{center}
    \includegraphics[width=0.9\hsize]{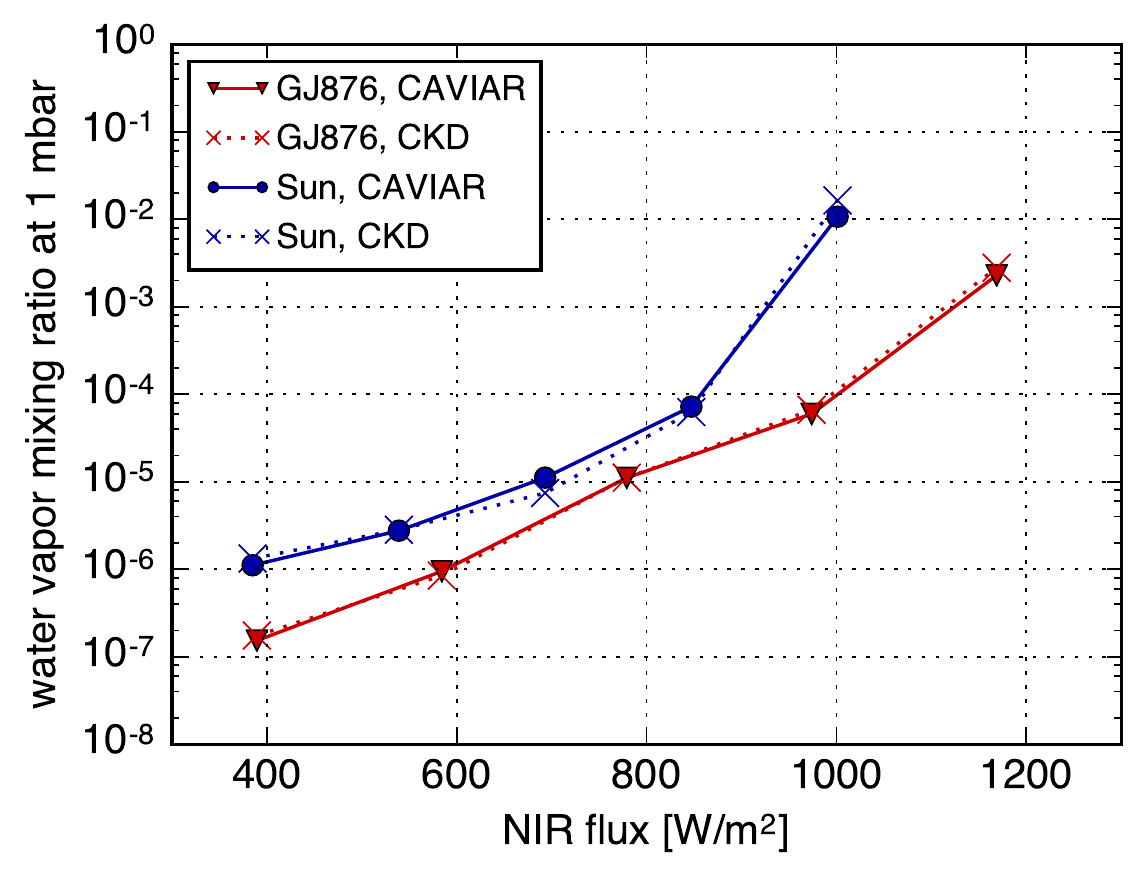}
    \end{center}
\caption{Effect of different water vapor continuum models on the water vapor mixing ratio at 1.39 mbar. Results with the CKD 2.4 continuum model, shown in dashed lines, are compared with our fiducial results with CAVIAR (same as the ones in Figure 3), in the case of the GJ~876 spectrum (red) and for the solar spectrum (blue). }
\label{fig:xH2O_CAVIAR_vs_CKD}
\end{figure}

\begin{figure}[!h]
    \begin{center}
    \includegraphics[width=0.9\hsize]{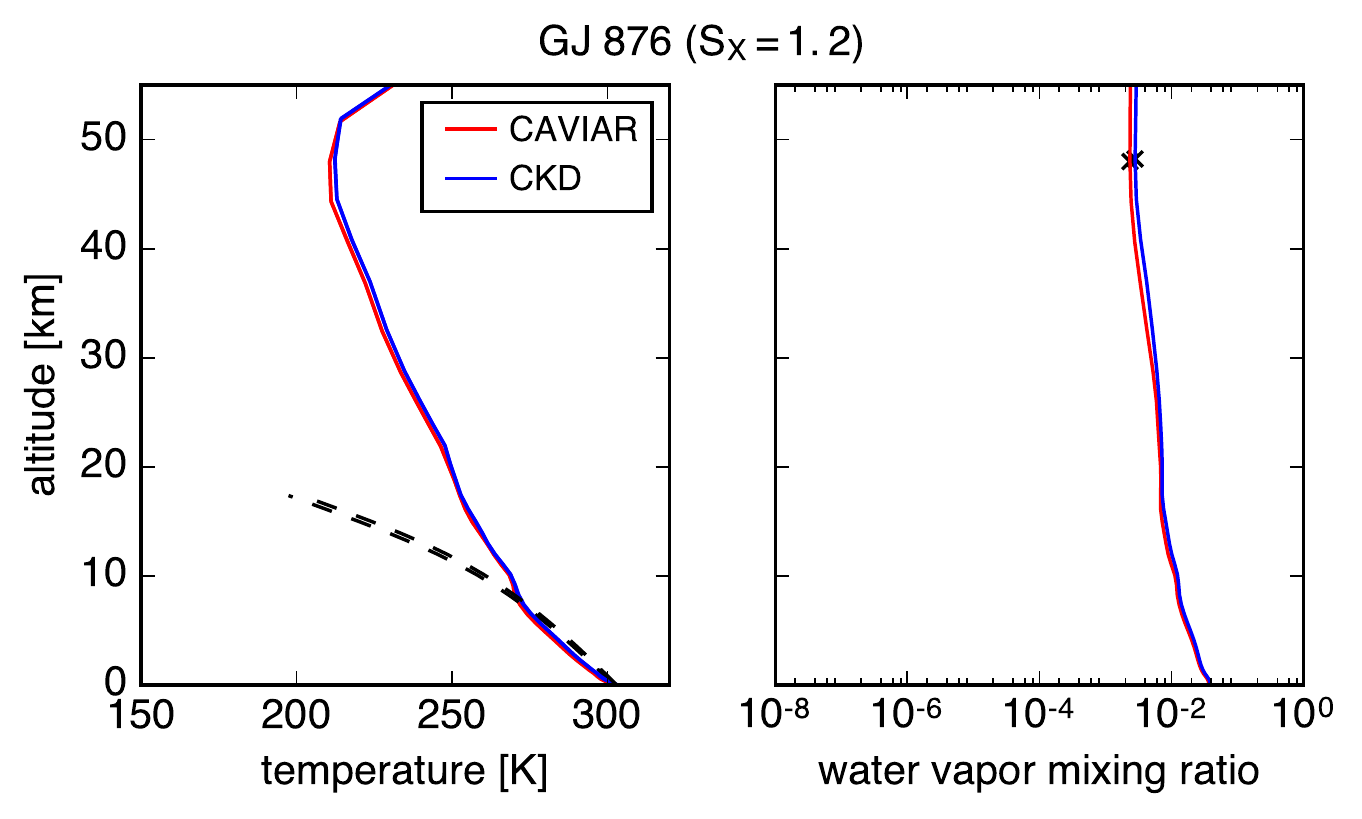}
    \includegraphics[width=0.9\hsize]{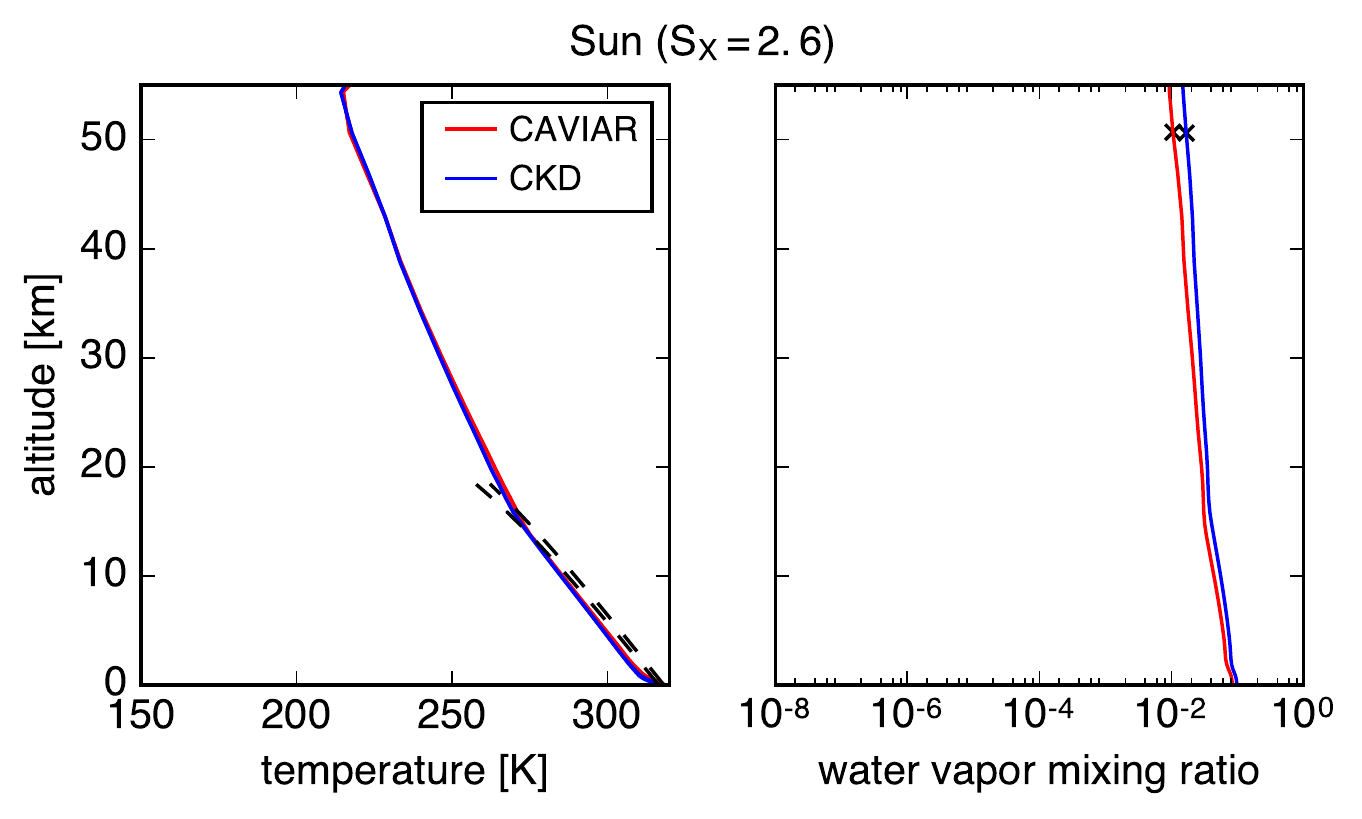}
    \end{center}
\caption{Effect of different water vapor continuum models on the vertical profiles of temperature and water vapor mixing ratio. Upper panels: GJ~876 spectrum and $S_X=1.2$. Lower panels: solar spectrum and $S_X=2.6$.}
\label{fig:CAVIAR_vs_CKD_GJ876_Sun}
\end{figure}

Figure \ref{fig:xH2O_CAVIAR_vs_CKD} compares \xwater{} with the CAVIAR model (our fiducial model) and those with the CKD 2.4 model, indicating that the effect of the water vapor continuum is very small. 
We also compare in Figure \ref{fig:CAVIAR_vs_CKD_GJ876_Sun} the vertical profiles of temperature and water vapor in the substellar region for the most highly irradiated (i.e., most moist) runs with the GJ~876 and Solar spectra. 
The resulting profiles overall closely match. 

When the atmospheric structure is fixed, the difference in short-wave fluxes due to the different water vapor continua is observed predominantly in the lower atmosphere ($<$ 10~km), as expected since the strength of continuum absorption is $\propto P^2$, where $P$ is the pressure. 
Using the $P$-$T$ and specific humidity profiles at the substellar point from our fiducial run with the spectrum of GJ~876 and $S_X = 1.2$, we find that the net short-wave clear-sky flux at the surface is larger with the CKD continuum compared to the CAVIAR continuum by about 7\%, with the net surface clear-sky short-wave flux with the CAVIAR model being 838 W/m$^2$. 
However, letting the system evolve, we find that such a difference is adjusted by small changes in temperature and humidity profiles and cloud patterns, leaving the basic atmospheric structure similar. 

Based on these results, we are confident that the correlation between the humidity in the upper atmosphere and the amount of NIR incident radiation shown in Figure \ref{fig:xH2O_S0X} is not sensitive to the choice of water vapor continuum model.

\section{Summary}
\label{s:summary}

We have investigated the response of atmospheric \wv{} profiles of synchronously rotating aquaplanets to changes in the incident stellar flux. 
We find a quasi-exponential increase of \wv{} mixing ratios in the upper atmosphere as a function of total incident flux, resulting in a variation of several orders of magnitude as the total incident flux increases by a factor of two to three, but this is much more gradual than what is predicted by 1D models. 

The increase of \wv{} mixing ratio in response to larger incident flux is not primarily caused by the increased surface temperature as previous 1D models suggest, due to the relatively stable surface temperature at the substellar point. 
Instead, it is induced by the effect of the increased NIR portion of the incident flux on the high-altitude circulation. 
The enhanced radiative heating due to the absorption by \wv{} and cloud particles drives a large-scale circulation of the air at altitudes above those to which convection penetrates, taking over the role of transporting \wv{} from the moist lower atmosphere to the upper atmosphere, and further increasing the heating rate in the upper atmosphere.
As a result, the moisture in the upper atmosphere increases dramatically,  while the surface below it remains temperate. 
This picture is supported by the good correlation between the \wv{} mixing ratio at \preslevel{} and the NIR portion of the incident radiation, regardless of the spectral type of the host star. 

Simulating transmission spectra based on the atmospheric profiles found in GCM experiments, 
we demonstrate that the increase of the \wv{} mixing ratio at \preslevel{} from $10^{-6}$ to $3 \times 10^{-3}$ leads to an increase of the \water{} absorption depth by a factor of a few, making the effective altitude of the \water{} absorption features as high as 40~km. 
We also find the presence of high clouds at the terminator, which    potentially increases the baseline effective altitude and thus weakens the absorption features. 
Our calculation may overestimate the effects of clouds on the transmission spectra due to the spatially and temporally averaged cloud properties of individual GCM grid cells.
Additional uncertainties come from other assumptions we made. 
A more realistic inclusion of clouds in our simulated transmission spectra will be the scope of future work. 

While our fiducial model assumed a thermodynamic ocean with zero ocean heat transport, 
comparable experiments with a dynamic ocean are also performed to demonstrate that this approximation of an ocean has only a modest effect on the \wv{} mixing ratio, the size of which is within about an order of magnitude or less. 
Varying the CO$_2$ mixing ratio from 1~ppm to 300~ppm has a negligible effect. 
We also confirmed that changing the planet spin period only has a secondary effect on the \wv{} mixing ratio, resulting in an order-of-magnitude variation at most. 
Our results are not sensitive to the choice of water vapor continuum  model within the known uncertainty. 

In this paper, we limit our discussion on the \wv{} profiles to the effects of the total incident flux and spectral energy distribution (i.e., the spectral type of the host star). 
These parameters are among the limited number of properties that can be constrained relatively easily by observations. 
In principle, however, \wv{} distributions also depend on geophysical parameters including the surface gravity, atmospheric composition, atmospheric pressure \citep[e.g.,][]{Turbet2016}, and the planetary water abundance, as well as photochemistry, which depends on the detailed atmospheric profiles. 
The dependence of the \wv{} mixing ratio on these planetary properties will be  the focus of future study.

\acknowledgments
We thank Michael J. Way and Nancy Y. Kiang for providing the boundary conditions for \modelE{} and for the helpful discussions. 
The support from Maxwell Kelley and Igor Aleinov was critical in resolving the issues identified in the process of generalizing \modelE{} for the exotic configurations considered in this paper. 
We thank James Manners for providing access to SOCRATES and advice on coupling it to \modelE{}. 
We also thank Philip Von Paris for the Kepler-186 spectrum. 
Y.~F. is deeply grateful to David S. Spiegel for the insightful discussions on the simulation suite for transmission spectra. 
Y.~F.'s research was supported by an appointment to the NASA Postdoctoral Program at the NASA Goddard Institute for Space Studies, administered by Universities Space Research Association under contract with NASA, and the Grant-in-Aid No.15K17605 by the Japan Society for the Promotion of Science. 
We also acknowledge support from the NASA Astrobiology Program through the Nexus for Exoplanet System Science.

\bibliography{GCM_qstrato_ref}

\begin{thebibliography}{}
\expandafter\ifx\csname natexlab\endcsname\relax\def\natexlab#1{#1}\fi

\bibitem[{{Abe} {et~al.}(2011){Abe}, {Abe-Ouchi}, {Sleep}, \&
  {Zahnle}}]{Abe2011}
{Abe}, Y., {Abe-Ouchi}, A., {Sleep}, N.~H., \& {Zahnle}, K.~J. 2011,
  Astrobiology, 11, 443

\bibitem[{{Allard} {et~al.}(2012){Allard}, {Homeier}, \&
  {Freytag}}]{Allard2012}
{Allard}, F., {Homeier}, D., \& {Freytag}, B. 2012, Philosophical Transactions
  of the Royal Society of London Series A, 370, 2765

\bibitem[{{Amundsen} {et~al.}(2017){Amundsen}, {Tremblin}, {Manners},
  {Baraffe}, \& {Mayne}}]{Amundsen2016}
{Amundsen}, D.~S., {Tremblin}, P., {Manners}, J., {Baraffe}, I., \& {Mayne},
  N.~J. 2017, \aap, 598, A97

\bibitem[{{Baines} \& {Armstrong}(2012)}]{Baines2012}
{Baines}, E.~K., \& {Armstrong}, J.~T. 2012, \apj, 744, 138

\bibitem[{{Baran} {et~al.}(2001){Baran}, {Francis}, {Labonnote}, \&
  {Doutriaux-Boucher}}]{Baran2001}
{Baran}, A.~J., {Francis}, P.~N., {Labonnote}, L.-C., \& {Doutriaux-Boucher},
  M. 2001, Quarterly Journal of the Royal Meteorological Society, 127, 2395

\bibitem[{{B{\'e}tr{\'e}mieux} \& {Kaltenegger}(2013)}]{Betremieux2013}
{B{\'e}tr{\'e}mieux}, Y., \& {Kaltenegger}, L. 2013, \apjl, 772, L31

\bibitem[{{B{\'e}tr{\'e}mieux} \& {Kaltenegger}(2014)}]{Betremieux2014}
---. 2014, \apj, 791, 7

\bibitem[{{Campargue} {et~al.}(2016){Campargue}, {Kassi}, {Mondelain},
  {Vasilchenko}, \& {Romanini}}]{Campargue2016}
{Campargue}, A., {Kassi}, S., {Mondelain}, D., {Vasilchenko}, S., \&
  {Romanini}, D. 2016, Journal of Geophysical Research (Atmospheres), 121, 13

\bibitem[{{Clough} {et~al.}(1989){Clough}, {Kneizys}, \& {Davies}}]{Clough1989}
{Clough}, S.~A., {Kneizys}, F.~X., \& {Davies}, R.~W. 1989, Atmospheric
  Research, 23, 229

\bibitem[{{Cowan} {et~al.}(2015){Cowan}, {Greene}, {Angerhausen}, {Batalha},
  {Clampin}, {Col{\'o}n}, {Crossfield}, {Fortney}, {Gaudi}, {Harrington},
  {Iro}, {Lillie}, {Linsky}, {Lopez-Morales}, {Mandell}, \&
  {Stevenson}}]{Cowan2015}
{Cowan}, N.~B., {Greene}, T., {Angerhausen}, D., {et~al.} 2015, \pasp, 127, 311

\bibitem[{{Del Genio} {et~al.}(2005){Del Genio}, {Kovari}, {Yao}, \&
  {Jonas}}]{DelGenio2005}
{Del Genio}, A.~D., {Kovari}, W., {Yao}, M.-S., \& {Jonas}, J. 2005, Journal of
  Climate, 18, 2376

\bibitem[{{Del Genio} {et~al.}(2007){Del Genio}, {Yao}, \&
  {Jonas}}]{DelGenio2007}
{Del Genio}, A.~D., {Yao}, M.-S., \& {Jonas}, J. 2007, \grl, 34, L16703

\bibitem[{{Del Genio} {et~al.}(1996){Del Genio}, {Yao}, {Kovari}, \&
  {Lo}}]{DelGenio1996}
{Del Genio}, A.~D., {Yao}, M.-S., {Kovari}, W., \& {Lo}, K.~K.-W. 1996, Journal
  of Climate, 9, 270

\bibitem[{{Dole}(1964)}]{Dole1964}
{Dole}, S.~H. 1964, {Habitable planets for man}

\bibitem[{{Domagal-Goldman} {et~al.}(2014){Domagal-Goldman}, {Segura},
  {Claire}, {Robinson}, \& {Meadows}}]{Domagal-Goldman2014}
{Domagal-Goldman}, S.~D., {Segura}, A., {Claire}, M.~W., {Robinson}, T.~D., \&
  {Meadows}, V.~S. 2014, \apj, 792, 90

\bibitem[{{Edson} {et~al.}(2011){Edson}, {Lee}, {Bannon}, {Kasting}, \&
  {Pollard}}]{Edson2011}
{Edson}, A., {Lee}, S., {Bannon}, P., {Kasting}, J.~F., \& {Pollard}, D. 2011,
  \icarus, 212, 1

\bibitem[{{Edwards}(1996)}]{Edwards1996}
{Edwards}, J.~M. 1996, Journal of Atmospheric Sciences, 53, 1921

\bibitem[{{Edwards} {et~al.}(2007){Edwards}, {Havemann}, {Thelen}, \&
  {Baran}}]{Edwards2007}
{Edwards}, J.~M., {Havemann}, S., {Thelen}, J.-C., \& {Baran}, A.~J. 2007,
  Atmospheric Research, 83, 19

\bibitem[{{Edwards} \& {Slingo}(1996)}]{EdwardsSlingo1996}
{Edwards}, J.~M., \& {Slingo}, A. 1996, Quarterly Journal of the Royal
  Meteorological Society, 122, 689

\bibitem[{{Ehrenreich} {et~al.}(2006){Ehrenreich}, {Tinetti}, {Lecavelier Des
  Etangs}, {Vidal-Madjar}, \& {Selsis}}]{Ehrenreich2006}
{Ehrenreich}, D., {Tinetti}, G., {Lecavelier Des Etangs}, A., {Vidal-Madjar},
  A., \& {Selsis}, F. 2006, \aap, 448, 379

\bibitem[{{Goldreich} \& {Peale}(1966)}]{Goldreich1966}
{Goldreich}, P., \& {Peale}, S. 1966, \aj, 71, 425

\bibitem[{{Goody} {et~al.}(1989){Goody}, {West}, {Chen}, \&
  {Crisp}}]{Goody1989}
{Goody}, R., {West}, R., {Chen}, L., \& {Crisp}, D. 1989, \jqsrt, 42, 539

\bibitem[{{Gregory}(2001)}]{Gregory2001}
{Gregory}, D. 2001, Quarterly Journal of the Royal Meteorological Society, 127,
  53

\bibitem[{{Hedelt} {et~al.}(2013){Hedelt}, {von Paris}, {Godolt}, {Gebauer},
  {Grenfell}, {Rauer}, {Schreier}, {Selsis}, \& {Trautmann}}]{Hedelt2013}
{Hedelt}, P., {von Paris}, P., {Godolt}, M., {et~al.} 2013, \aap, 553, A9

\bibitem[{{Ishiwatari} {et~al.}(2002){Ishiwatari}, {Takehiro}, {Nakajima}, \&
  {Hayashi}}]{Ishiwatari2002}
{Ishiwatari}, M., {Takehiro}, S.-I., {Nakajima}, K., \& {Hayashi}, Y.-Y. 2002,
  Journal of Atmospheric Sciences, 59, 3223

\bibitem[{{Kaltenegger} \& {Traub}(2009)}]{Kaltenegger2009}
{Kaltenegger}, L., \& {Traub}, W.~A. 2009, \apj, 698, 519

\bibitem[{{Kasting} {et~al.}(2015){Kasting}, {Chen}, \&
  {Kopparapu}}]{Kasting2015}
{Kasting}, J.~F., {Chen}, H., \& {Kopparapu}, R.~K. 2015, \apjl, 813, L3

\bibitem[{{Kasting} {et~al.}(1993){Kasting}, {Whitmire}, \&
  {Reynolds}}]{Kasting1993}
{Kasting}, J.~F., {Whitmire}, D.~P., \& {Reynolds}, R.~T. 1993, \icarus, 101,
  108

\bibitem[{{Kodama} {et~al.}(2015){Kodama}, {Genda}, {Abe}, \&
  {Zahnle}}]{Kodama2015}
{Kodama}, T., {Genda}, H., {Abe}, Y., \& {Zahnle}, K.~J. 2015, \apj, 812, 165

\bibitem[{{Kopparapu} {et~al.}(2016){Kopparapu}, {Wolf}, {Haqq-Misra}, {Yang},
  {Kasting}, {Meadows}, {Terrien}, \& {Mahadevan}}]{Kopparapu2016}
{Kopparapu}, R.~k., {Wolf}, E.~T., {Haqq-Misra}, J., {et~al.} 2016, \apj, 819,
  84

\bibitem[{{Kopparapu} {et~al.}(2013){Kopparapu}, {Ramirez}, {Kasting}, {Eymet},
  {Robinson}, {Mahadevan}, {Terrien}, {Domagal-Goldman}, {Meadows}, \&
  {Deshpande}}]{Kopparapu2013}
{Kopparapu}, R.~K., {Ramirez}, R., {Kasting}, J.~F., {et~al.} 2013, \apj, 765,
  131

\bibitem[{{Lacis} \& {Oinas}(1991)}]{Lacis1991}
{Lacis}, A.~A., \& {Oinas}, V. 1991, \jgr, 96, 9027

\bibitem[{{Lean} {et~al.}(2005){Lean}, {Rottman}, {Harder}, \&
  {Kopp}}]{Lean2005}
{Lean}, J., {Rottman}, G., {Harder}, J., \& {Kopp}, G. 2005, \solphys, 230, 27

\bibitem[{{Leconte} {et~al.}(2013{\natexlab{a}}){Leconte}, {Forget}, {Charnay},
  {Wordsworth}, \& {Pottier}}]{Leconte2013a}
{Leconte}, J., {Forget}, F., {Charnay}, B., {Wordsworth}, R., \& {Pottier}, A.
  2013{\natexlab{a}}, \nat, 504, 268

\bibitem[{{Leconte} {et~al.}(2013{\natexlab{b}}){Leconte}, {Forget}, {Charnay},
  {Wordsworth}, {Selsis}, {Millour}, \& {Spiga}}]{Leconte2013b}
{Leconte}, J., {Forget}, F., {Charnay}, B., {et~al.} 2013{\natexlab{b}}, \aap,
  554, A69

\bibitem[{{Leconte} {et~al.}(2015){Leconte}, {Wu}, {Menou}, \&
  {Murray}}]{Leconte2015}
{Leconte}, J., {Wu}, H., {Menou}, K., \& {Murray}, N. 2015, Science, 347, 632

\bibitem[{{Merlis} \& {Schneider}(2010)}]{Merlis2010}
{Merlis}, T.~M., \& {Schneider}, T. 2010, Journal of Advances in Modeling Earth
  Systems, 2, 13

\bibitem[{{Misra} {et~al.}(2014){Misra}, {Meadows}, \& {Crisp}}]{Misra2014}
{Misra}, A., {Meadows}, V., \& {Crisp}, D. 2014, \apj, 792, 61

\bibitem[{{Mlawer} {et~al.}(2012){Mlawer}, {Payne}, {Moncet}, {Delamere},
  {Alvarado}, \& {Tobin}}]{Mlawer2012}
{Mlawer}, E.~J., {Payne}, V.~H., {Moncet}, J.-L., {et~al.} 2012, Philosophical
  Transactions of the Royal Society of London Series A, 370, 2520

\bibitem[{{Ptashnik} {et~al.}(2011){Ptashnik}, {McPheat}, {Shine}, {Smith}, \&
  {Williams}}]{Ptashnik2011}
{Ptashnik}, I.~V., {McPheat}, R.~A., {Shine}, K.~P., {Smith}, K.~M., \&
  {Williams}, R.~G. 2011, Journal of Geophysical Research (Atmospheres), 116,
  D16305

\bibitem[{{Quintana} {et~al.}(2014){Quintana}, {Barclay}, {Raymond}, {Rowe},
  {Bolmont}, {Caldwell}, {Howell}, {Kane}, {Huber}, {Crepp}, {Lissauer},
  {Ciardi}, {Coughlin}, {Everett}, {Henze}, {Horch}, {Isaacson}, {Ford},
  {Adams}, {Still}, {Hunter}, {Quarles}, \& {Selsis}}]{Quintana2014}
{Quintana}, E.~V., {Barclay}, T., {Raymond}, S.~N., {et~al.} 2014, Science,
  344, 277

\bibitem[{{Rauer} {et~al.}(2011){Rauer}, {Gebauer}, {Paris}, {Cabrera},
  {Godolt}, {Grenfell}, {Belu}, {Selsis}, {Hedelt}, \& {Schreier}}]{Rauer2011}
{Rauer}, H., {Gebauer}, S., {Paris}, P.~V., {et~al.} 2011, \aap, 529, A8

\bibitem[{{Rothman} {et~al.}(2013){Rothman}, {Gordon}, {Babikov}, {Barbe},
  {Chris Benner}, {Bernath}, {Birk}, {Bizzocchi}, {Boudon}, {Brown},
  {Campargue}, {Chance}, {Cohen}, {Coudert}, {Devi}, {Drouin}, {Fayt}, {Flaud},
  {Gamache}, {Harrison}, {Hartmann}, {Hill}, {Hodges}, {Jacquemart}, {Jolly},
  {Lamouroux}, {Le Roy}, {Li}, {Long}, {Lyulin}, {Mackie}, {Massie},
  {Mikhailenko}, {M{\"u}ller}, {Naumenko}, {Nikitin}, {Orphal}, {Perevalov},
  {Perrin}, {Polovtseva}, {Richard}, {Smith}, {Starikova}, {Sung}, {Tashkun},
  {Tennyson}, {Toon}, {Tyuterev}, \& {Wagner}}]{Rothman2013}
{Rothman}, L.~S., {Gordon}, I.~E., {Babikov}, Y., {et~al.} 2013, \jqsrt, 130, 4

\bibitem[{{Rugheimer} {et~al.}(2015){Rugheimer}, {Kaltenegger}, {Segura},
  {Linsky}, \& {Mohanty}}]{Rugheimer2015}
{Rugheimer}, S., {Kaltenegger}, L., {Segura}, A., {Linsky}, J., \& {Mohanty},
  S. 2015, \apj, 809, 57

\bibitem[{{Rugheimer} {et~al.}(2013){Rugheimer}, {Kaltenegger}, {Zsom},
  {Segura}, \& {Sasselov}}]{Rugheimer2013}
{Rugheimer}, S., {Kaltenegger}, L., {Zsom}, A., {Segura}, A., \& {Sasselov}, D.
  2013, Astrobiology, 13, 251

\bibitem[{{Schmidt} {et~al.}(2014){Schmidt}, {Kelley}, {Nazarenko}, {Ruedy},
  {Russell}, {Aleinov}, {Bauer}, {Bauer}, {Bhat}, {Bleck}, {Canuto}, {Chen},
  {Cheng}, {Clune}, {Del Genio}, {de Fainchtein}, {Faluvegi}, {Hansen},
  {Healy}, {Kiang}, {Koch}, {Lacis}, {LeGrande}, {Lerner}, {Lo}, {Matthews},
  {Menon}, {Miller}, {Oinas}, {Oloso}, {Perlwitz}, {Puma}, {Putman}, {Rind},
  {Romanou}, {Sato}, {Shindell}, {Sun}, {Syed}, {Tausnev}, {Tsigaridis},
  {Unger}, {Voulgarakis}, {Yao}, \& {Zhang}}]{Schmidt2014}
{Schmidt}, G.~A., {Kelley}, M., {Nazarenko}, L., {et~al.} 2014, Journal of
  Advances in Modeling Earth Systems, 6, 141

\bibitem[{{Segura} {et~al.}(2005){Segura}, {Kasting}, {Meadows}, {Cohen},
  {Scalo}, {Crisp}, {Butler}, \& {Tinetti}}]{Segura2005}
{Segura}, A., {Kasting}, J.~F., {Meadows}, V., {et~al.} 2005, Astrobiology, 5,
  706

\bibitem[{{Segura} {et~al.}(2003){Segura}, {Krelove}, {Kasting}, {Sommerlatt},
  {Meadows}, {Crisp}, {Cohen}, \& {Mlawer}}]{Segura2003}
{Segura}, A., {Krelove}, K., {Kasting}, J.~F., {et~al.} 2003, Astrobiology, 3,
  689

\bibitem[{{Shine} {et~al.}(2016){Shine}, {Campargue}, {Mondelain}, {McPheat},
  {Ptashnik}, \& {Weidmann}}]{Shine2016}
{Shine}, K.~P., {Campargue}, A., {Mondelain}, D., {et~al.} 2016, Journal of
  Molecular Spectroscopy, 327, 193

\bibitem[{{Sing} {et~al.}(2016){Sing}, {Fortney}, {Nikolov}, {Wakeford},
  {Kataria}, {Evans}, {Aigrain}, {Ballester}, {Burrows}, {Deming},
  {D{\'e}sert}, {Gibson}, {Henry}, {Huitson}, {Knutson}, {Lecavelier Des
  Etangs}, {Pont}, {Showman}, {Vidal-Madjar}, {Williamson}, \&
  {Wilson}}]{Sing2016}
{Sing}, D.~K., {Fortney}, J.~J., {Nikolov}, N., {et~al.} 2016, \nat, 529, 59

\bibitem[{{Tinetti} {et~al.}(2007){Tinetti}, {Vidal-Madjar}, {Liang},
  {Beaulieu}, {Yung}, {Carey}, {Barber}, {Tennyson}, {Ribas}, {Allard},
  {Ballester}, {Sing}, \& {Selsis}}]{Tinetti2007}
{Tinetti}, G., {Vidal-Madjar}, A., {Liang}, M.-C., {et~al.} 2007, \nat, 448,
  169

\bibitem[{{Torres} {et~al.}(2015){Torres}, {Kipping}, {Fressin}, {Caldwell},
  {Twicken}, {Ballard}, {Batalha}, {Bryson}, {Ciardi}, {Henze}, {Howell},
  {Isaacson}, {Jenkins}, {Muirhead}, {Newton}, {Petigura}, {Barclay},
  {Borucki}, {Crepp}, {Everett}, {Horch}, {Howard}, {Kolbl}, {Marcy},
  {McCauliff}, \& {Quintana}}]{Torres2015}
{Torres}, G., {Kipping}, D.~M., {Fressin}, F., {et~al.} 2015, \apj, 800, 99

\bibitem[{{Turbet} {et~al.}(2016){Turbet}, {Leconte}, {Selsis}, {Bolmont},
  {Forget}, {Ribas}, {Raymond}, \& {Anglada-Escud{\'e}}}]{Turbet2016}
{Turbet}, M., {Leconte}, J., {Selsis}, F., {et~al.} 2016, \aap, 596, A112

\bibitem[{{van der Werf}(2008)}]{vanderWerf2008}
{van der Werf}, S.~Y. 2008, \ao, 47, 153

\bibitem[{{von Braun} {et~al.}(2014){von Braun}, {Boyajian}, {van Belle},
  {Kane}, {Jones}, {Farrington}, {Schaefer}, {Vargas}, {Scott}, {ten
  Brummelaar}, {Kephart}, {Gies}, {Ciardi}, {L{\'o}pez-Morales}, {Mazingue},
  {McAlister}, {Ridgway}, {Goldfinger}, {Turner}, \& {Sturmann}}]{vonBraun2014}
{von Braun}, K., {Boyajian}, T.~S., {van Belle}, G.~T., {et~al.} 2014, \mnras,
  438, 2413

\bibitem[{{Way} {et~al.}(2015){Way}, {Del Genio}, {Kelley}, {Aleinov}, \&
  {Clune}}]{Way2015}
{Way}, M.~J., {Del Genio}, A.~D., {Kelley}, M., {Aleinov}, I., \& {Clune}, T.
  2015, ArXiv e-prints, arXiv:1511.07283

\bibitem[{{Way} {et~al.}(2016){Way}, {Del Genio}, {Kiang}, {Sohl}, {Grinspoon},
  {Aleinov}, {Kelley}, \& {Clune}}]{Way2016}
{Way}, M.~J., {Del Genio}, A.~D., {Kiang}, N.~Y., {et~al.} 2016, \grl, 43, 8376

\bibitem[{{Way} {et~al.}(2017){Way}, {Aleinov}, {Amundsen}, {Chandler},
  {Clune}, {Del Genio}, {Fujii}, {Kelley}, {Kiang}, {Sohl}, \&
  {Tsigaridis}}]{Way2017}
{Way}, M.~J., {Aleinov}, I., {Amundsen}, D.~S., {et~al.} 2017, \apjs, 231, 12

\bibitem[{{Wiscombe} \& {Evans}(1977)}]{Wiscombe1977}
{Wiscombe}, W.~J., \& {Evans}, J.~W. 1977, Journal of Computational Physics,
  24, 416

\bibitem[{{Wolf} \& {Toon}(2014)}]{Wolf2014}
{Wolf}, E.~T., \& {Toon}, O.~B. 2014, \grl, 41, 167

\bibitem[{{Wolf} \& {Toon}(2015)}]{Wolf2015}
---. 2015, Journal of Geophysical Research (Atmospheres), 120, 5775

\bibitem[{{Wordsworth} \& {Pierrehumbert}(2014)}]{Wordsworth2014}
{Wordsworth}, R., \& {Pierrehumbert}, R. 2014, \apjl, 785, L20

\bibitem[{{Wordsworth} \& {Pierrehumbert}(2013)}]{Wordsworth2013}
{Wordsworth}, R.~D., \& {Pierrehumbert}, R.~T. 2013, \apj, 778, 154

\bibitem[{{Yang} {et~al.}(2014){Yang}, {Bou{\'e}}, {Fabrycky}, \&
  {Abbot}}]{Yang2014}
{Yang}, J., {Bou{\'e}}, G., {Fabrycky}, D.~C., \& {Abbot}, D.~S. 2014, \apjl,
  787, L2

\bibitem[{{Yang} {et~al.}(2013){Yang}, {Cowan}, \& {Abbot}}]{Yang2013}
{Yang}, J., {Cowan}, N.~B., \& {Abbot}, D.~S. 2013, \apjl, 771, L45

\bibitem[{{Zdunkowski} \& {Korb}(1985)}]{Zdunkowski1985}
{Zdunkowski}, W.~G., \& {Korb}, G.~J. 1985, Promet, 2/3, 26

\bibitem[{{Zdunkowski} {et~al.}(1980){Zdunkowski}, {Welch}, \&
  {Korb}}]{Zdunkowski1980}
{Zdunkowski}, W.~G., {Welch}, R.~M., \& {Korb}, G. 1980, Beitr\"{a}ge zur
  Physik der Atmosph\"{a}re, 53, 147

\end{thebibliography}

\appendix

\section{Validation of radiation scheme\\used in \modelE{} }
\label{ap:radiation}

In this paper we applied the ROCKE-3D GCM to planets around stars with different spectral types. 
In doing so, special caution is needed regarding the accuracy of the short-wave radiative transfer calculation, because the baseline model of ROCKE-3D was  developed for the Earth, and the radiative transfer calculation was optimized for the solar spectrum. 
We therefore modified the bands used by the short-wave radiation calculation in SOCRATES in order to improve accuracy for low-mass stars while at the same time maintaining the accuracy for Sun-like stars.
Our adopted short-wave bands in this study are given in Table \ref{tab:short-wave}.

In order to check the accuracy of the short-wave radiative transfer calculation used in the GCM, we compare it to a more accurate higher-resolution (280 bands) calculation using the same atmospheric profile. 
We use the atmospheric profile at the substellar point of a planet around GJ~876 with $S_X=1.2$ (see Figure~\ref{fig:AqOH0TLS_GJ876_temp_xH2O_vz_heat}), but without clouds. 
This profile was chosen because of its extremely high \wv{} mixing ratio, which is very different from that of the Earth. 
Assuming this profile, we calculated short-wave radiative fluxes with 29 bands (used in the GCM) and with 280 bands, with both the Sun and GJ~876 as the incident spectra.

\begin{figure*}[!htb]
    \begin{center}
    \includegraphics[width=0.8\hsize]{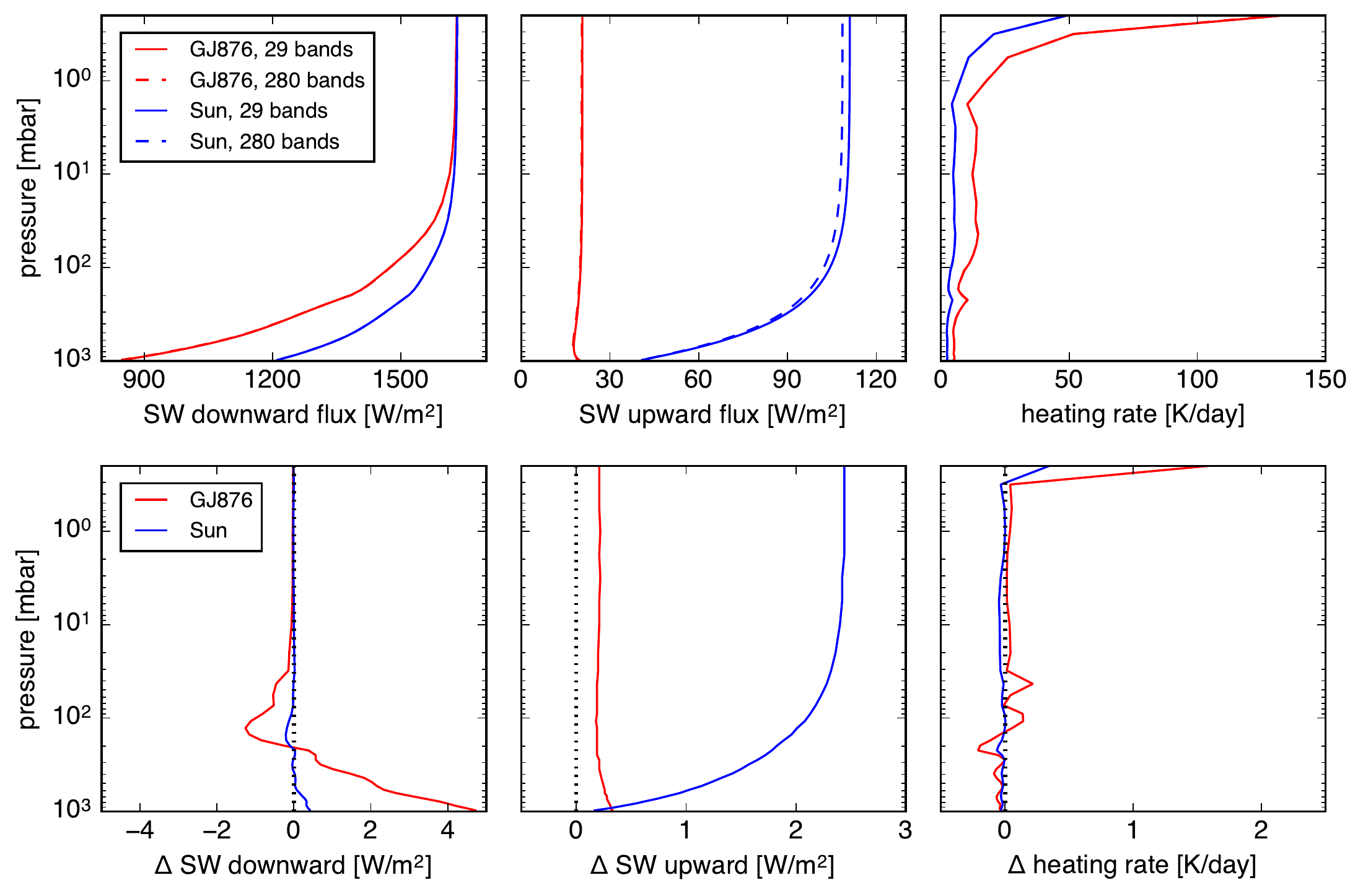}
    \end{center}
\caption{Comparison of short-wave downward fluxes (left), upward fluxes (center), and heating rates (right) between the 29-band short-wave radiative transfer used in the GCM and the 280-band benchmark. The atmospheric profile assumed is the one at the substellar point of a planet around GJ 876 with $S_X=1.2$ (see the upper panels of Figure~\ref{fig:AqOH0TLS_GJ876_temp_xH2O_vz_heat}). }
\label{fig:socrates}
\end{figure*}

We show in Figure~\ref{fig:socrates} the short-wave (stellar) downward fluxes, upward fluxes and heating rates we obtained.
Differences between the 29- and 280-band configurations are small, $<$ 5 W/m$^2$ for fluxes and $<$ 0.5 K/day for heating rates except at pressures $\ll$ 1 mbar, throughout the atmosphere. We therefore conclude that the radiation scheme adopted here is sufficiently accurate.

\begin{table}
\centering
\caption{The short-wave bands.}
\label{tab:short-wave}
\begin{tabular}{l|l|l} \hline \hline
\# & lower limit [\textmu m] & upper limit [\textmu m] \\ \hline 
    1 &       0.200 &    0.385 \\
    2 &       0.385 &    0.500 \\
    3 &       0.500 &    0.690 \\
    4 &       0.690 &    0.870 \\
    5 &       0.870 &    0.900 \\
    6 &       0.900 &    1.08 \\
    7 &       1.08  &   1.12 \\
    8 &       1.12  &   1.16 \\
    9 &       1.16  &   1.20 \\
   10 &       1.20  &   1.30 \\
   11 &       1.30  &   1.34 \\
   12 &       1.34  &   1.42 \\
   13 &       1.42  &   1.46 \\
   14 &       1.46  &   1.52 \\
   15 &       1.52  &   1.56 \\
   16 &       1.56  &   1.62 \\
   17 &       1.62  &   1.68 \\
   18 &       1.68  &   1.80 \\
   19 &       1.80  &   1.94 \\
   20 &       1.94  &   2.00 \\
   21 &       2.00  &   2.14 \\
   22 &       2.14  &   2.50 \\
   23 &       2.50  &   2.65 \\
   24 &       2.65  &   2.85 \\
   25 &       2.85  &   3.15 \\
   26 &       3.15  &   3.60 \\
   27 &       3.60  &   4.10 \\
   28 &       4.10  &   4.60 \\
   29 &       4.60  &   10.0 \\ \hline
\end{tabular}
\end{table}

\end{document}